**Full Paper**

**Correlated In-Situ Low-Frequency Noise and Impedance Spectroscopy Reveal Recombination Dynamics in Organic Solar Cells using Fullerene and Non-Fullerene Acceptors**

*Kyle A. Luck, Vinod K. Sangwan, Patrick E. Hartnett, Heather N. Arnold, Michael R. Wasielewski, Tobin J. Marks, and Mark C. Hersam\**

K. A. Luck, Dr. V. K. Sangwan, Dr. H. N. Arnold, Prof. T. J. Marks, Prof. M. C. Hersam
Northwestern University, Department of Materials Science and Engineering, and the Argonne-Northwestern Solar Energy Research Center, 2220 Campus Drive, Evanston, IL 60208, USA
E-mail: m-hersam@northwestern.edu

P. E. Hartnett, Prof. M. R. Wasielewski, Prof. T. J. Marks, Prof. M. C. Hersam
Northwestern University, Department of Chemistry, and the Argonne-Northwestern Solar Energy Research Center, 2145 Sheridan Road, Evanston, IL 60208, USA

Prof. M. C. Hersam
Northwestern University, Department of Electrical Engineering and Computer Science, 2145 Sheridan Road, Evanston, IL 60208, USA

**Abstract**

Non-fullerene acceptors based on perylenediimides (PDIs) have garnered significant interest as an alternative to fullerene acceptors in organic photovoltaics (OPVs), but their charge transport phenomena are not well understood, especially in bulk heterojunctions (BHJs). Here, we investigate charge transport and current fluctuations by performing correlated low-frequency noise and impedance spectroscopy measurements on two BHJ OPV systems, one employing a fullerene acceptor and the other employing a dimeric PDI acceptor. In the dark, these measurements reveal that PDI-based OPVs have a greater degree of recombination in comparison to fullerene-based OPVs. Furthermore, for the first time in organic solar cells, 1/f noise data are fit to the Kleinpenning model to reveal underlying current fluctuations in different transport regimes. Under illumination, 1/f noise increases by approximately four orders of magnitude for the fullerene-based OPVs and three orders of magnitude for the PDI-based OPVs. An inverse correlation is also observed between noise spectral density and power conversion efficiency. Overall, these results show that low-frequency noise spectroscopy is an effective in-situ diagnostic tool to assess charge transport in emerging photovoltaic materials, thereby providing quantitative guidance for the design of next-generation solar cell materials and technologies.

## 1. Introduction

Following the seminal demonstration of bilayer heterojunction organic photovoltaic (OPV) cells,[1] intense research effort has been devoted toward accelerating the practical feasibility of OPVs. This research is driven in part by the earth abundance and synthetic tunability of organic semiconductor constituents, as well as the potential for low-cost, high-throughput manufacturing on flexible substrates.[2,3] Following the early work on bilayer OPVs, the next major advance occurred with the introduction of the bulk heterojunction (BHJ) architecture. By employing a solution-processable blend of a polymer donor and fullerene acceptor,[4] the BHJ architecture enables thicker photoactive layers to harvest more of the solar spectrum while maintaining efficient exciton dissociation throughout the entire photoactive blend.[5] Recently, fullerene-based BHJ cells have not only achieved single-junction power conversion efficiencies in excess of 10%,[6,7] but have also been integrated with large-area flexible substrates.[8–10] However, a key limitation of fullerene acceptors is their energy intensive synthesis and environmental instability[11,12] rendering them non-ideal for mass production.

The scalability limitation of fullerenes has motivated the exploration of high-performance non-fullerene acceptors.[13–18] Among the most widely studied classes of non-fullerene acceptors are perylenediimide (PDI) derivatives,[19–21] which are promising due to strong light absorption, high electron mobilities, environmental stability, and amenability to mass production.[22–26] However, the performance of OPVs employing PDI derivatives has lagged behind their fullerene counterparts[27] due to increased tendency to form large crystalline domains[28] that inhibit exciton splitting. Consequently, research efforts have been devoted to strategies to disrupt long-range PDI crystallinity including the incorporation of "swallowtail" side groups[29] and functionalizing the "headland" positions to achieve slip-stacking.[30] Additional work has focused on using fused PDI units including dimers,[31,32] trimers,[33] and tetramers,[34] which have attained performance rivaling fullerenes.

While these strategies have enabled BHJ morphologies with enhanced efficiency, PDIs still exhibit a significant barrier to charge separation.[35,36] Conversely, fullerenes exhibit essentially barrierless charge separation,[37] enabling ultrafast exciton dissociation with various donor materials such as poly(3-hexylthiophene-2,5-diyl) (P3HT) and poly[[4,8-bis[(2-ethylhexyl)oxy]benzo[1,2-b:4,5-b′]dithiophene-2,6-diyl][3-fluoro-2-[(2-ethylhexyl)carbonyl]-thieno[3,4-b]thiophenediyl]] (PTB7).[38–41] Following exciton dissociation, separated charges must travel through the remainder of the active layer to be collected at their respective electrodes. Charge transport can occur via either a drift-dominated or a diffusion-dominated mechanism depending on the operating regime and heterojunction materials, morphology, and processing conditions.[40,42,43] While preliminary theoretical work predicts less efficient charge transport in PDI networks in comparison to fullerene networks,[44] charge transport in BHJ OPVs employing PDI acceptors has not been thoroughly characterized experimentally.[45]

Low-frequency noise (LFN) spectroscopy is an in-situ, non-destructive technique to probe charge transport in OPVs that reveals information about recombination dynamics and fluctuations not accessible by standard current-voltage characteristics. Excess $1/f$ noise in the bandwidth of 1 Hz – 10 kHz (below white noise) originating from resistance fluctuations is particularly useful for elucidating trapping-detrapping and scattering processes.[46] LFN spectroscopy is a previously established diagnostic tool for silicon diode and photovoltaic reliability.[47–49] Furthermore, LFN has been used to more accurately characterize the specific detectivity of photodiodes and photodetectors, since photodetector performance is limited by both shot noise and excess $1/f$ noise.[50–54] Moreover, noise spectroscopy has been used to probe percolating conduction networks in organic blend films.[55] However, LFN measurements have been relatively infrequently utilized for OPVs, particularly for devices employing non-fullerene acceptors.

In contrast, other techniques, such as transient photovoltage, differential charging experiments, and impedance spectroscopy have been more widely used to probe OPV recombination dynamics.[56–59] With regard to impedance spectroscopy, significant effort has been devoted to develop robust equivalent circuit models incorporating both resistive and capacitive elements. These models, in conjunction with working hypotheses, enable the determination of relevant device parameters such as the recombination resistance, chemical capacitance, and charge carrier lifetimes that ultimately impact power conversion efficiency (PCE).[60] Impedance spectroscopy measurements have revealed solar cell charge recombination dynamics over a large frequency bandwidth (typically, 1 Hz – 1 MHz), in different bias regimes, and under varying illumination conditions.[60–64] Even materials parameters such as carrier mobility, carrier lifetime, and density of states have been extracted from impedance spectroscopy.[65] Overall, impedance spectroscopy and LFN measurements provide complementary perspectives on charge transport phenomena, suggesting that a combined approach will be particularly useful for emerging OPV and related photovoltaic systems.

Here, we study current fluctuations via LFN measurements and impedance spectroscopy in OPV BHJ systems utilizing poly[4,8-bis(5-(2-ethylhexyl)thiophen-2-yl)benzo[1,2-b;4,5-b$'$]dithiophene-2,6-diyl-alt-(4-(2-ethylhexyl)-3-fluorothieno[3,4-b]thiophene-)-2-carboxylate-2-6-diyl)] (PBDTTT-EFT) (Figure 1a)[66,67] as the electron donor and either [6,6]-phenyl-$C_{71}$-butyric acid methyl ester ($PC_{71}BM$)[66,68] or [1,2:3,4]-bis-[$N,N'$-bis-perylenediimide-1,12-yl]-benzene (Ph2a)[31] as the electron acceptor (Figure 1a). These BHJs enable comparisons between OPVs utilizing the recently discovered high-performance dimeric PDI acceptor, Ph2a, and a conventional fullerene acceptor ($PC_{71}BM$). We first analyze current-voltage characteristics in the $PBDTTT-EFT:PC_{71}BM$ and PBDTTT-EFT:Ph2a OPV systems. These direct current (DC) measurements confirm higher PCE for the $PBDTTT-EFT:PC_{71}BM$ system, as expected.[67] Next impedance spectroscopy and LFN spectroscopy measurements are performed under dark conditions over a range of biases to extract steady-state *p-n* heterojunction models. All devices show $1/f$ behavior at a sufficiently small frequency bandwidth where thermal and shot noise is minimized. These measurements reveal a narrower density-of-states (DOS) profile in the Ph2a-based cells in comparison to fullerene-based OPVs. A narrower DOS profile points to a higher degree of recombination in PBDTTT-EFT:Ph2a compared to $PBDTTT-EFT–PC_{71}BM$, as an inverse correlation between the proportion of charge traps (which lead to recombination) and the density-of-states breadth has been previously observed in semiconducting organic blend films.[39,55]

Furthermore, for the first time in OPVs, we fit bias-dependent LFN data to the Kleinpenning model.[69] This fit is made possible by extracting the carrier lifetime ($\tau$) by impedance spectroscopy. Subsequently, correlated LFN and impedance measurements are studied under 1 sun illumination. Noise spectral density is found to increase for both systems under illumination, with lower spectral density being correlated with higher PCE. Overall, this study establishes correlated $1/f$ noise and impedance spectroscopy as a powerful in-situ analytical tool for OPVs to probe recombination dynamics. This information can be used to inform materials selection and processing parameters, with the potential to streamline device optimization for emerging photovoltaic materials.

## 2. Results and Discussion

### 2.1. Solar Cell Performance

The chemical structures of the active layer materials used in this study, including PBDTTT-EFT, $PC_{71}BM$, and Ph2a, are shown in Figure 1a. The PBDTTT-EFT donor polymer is a synthetic extension of the well-known donor polymer (PTB7).[66,67] In PBDTTT-EFT, the 2-ethylhexyl-oxy side chains of PTB7 are substituted with 2-ethylhexyl-thienyl side chains, which concurrently extends absorption and increases the highest occupied molecular orbital (HOMO) energy. Among fullerene acceptors, $PC_{71}BM$ pairs particularly well with PTB7 and its derivatives due to its complementary solar spectrum absorption, favorable lowest unoccupied molecular orbital (LUMO) energy level alignment, and high carrier mobility, affording high internal quantum efficiencies (IQEs) in OPVs.[66,68] In addition, the high-lying HOMO energy in PBDTTT-EFT increases the HOMO-LUMO offset in PBDTTT-EFT:$PC_{71}BM$ OPVs, thereby increasing the device open-circuit voltage compared to that of the PTB7$PC_{71}BM$ system.[67,70]

While fullerenes have achieved high performance with PTB7-based donor materials, their energy-intensive synthesis has, as noted above, motivated exploration of non-fullerene acceptor materials,[11] especially PDI derivatives that can be manufactured at low cost.[30] Note that the PBDTTT-EFT donor polymer exhibits excellent performance in OPV cells that employ the dimeric PDI acceptor, Ph2a.[31] Ph2a exhibits complementary absorption with PBDTTT-EFT and its fused, twisted molecular structure translates into desired BHJ morphologies with high electron mobility.[31] In this work, the PBDTTT-EFT, $PC_{71}BM$, and Ph2a active layer materials are utilized in an inverted OPV BHJ architecture (Figure 1b). This device geometry employs commonly used interfacial layers, namely sol-gel zinc oxide (ZnO) as the electron transport layer and thermally evaporated molybdenum oxide ($MoO_x$) as the hole transport layer.[71] In this study, all device fabrication steps following the ZnO layer deposition are carried out in an argon-filled glovebox. Samples are encapsulated prior to removal from the glovebox and are stable over the timescale of current-voltage, LFN, and impedance spectroscopy measurements in ambient conditions.

Figure 1c shows the current density-voltage (*J-V*) characteristics of representative PBDTTT-EFT:$PC_{71}BM$ and PBDTTT-EFT:Ph2a devices in the dark. As has been observed previously, the *J-V* characteristics reveal higher leakage currents in the PBDTTT-EFT:Ph2a device than the PBDTTT-ETT:PC$_{71}$BM device.[44] To gain deeper physical insight into the *J-V* characteristics, these curves were fit under forward bias using the Shockley diode equation with a series resistor term (eq. 1),[72]

$$I = I_0 e^{\frac{q(V - IR_s)}{nk_BT}} \tag{1}$$

where $I$ is current, $I_0$ is reverse bias saturation current, $q$ is the elementary unit of charge, $V$ is the applied bias, $R_s$ is the series resistance, $n$ is the ideality factor ($n = 1$ for an ideal diode), $k_B$ is Boltzmann's constant, and $T$ is temperature. The fits are shown in Figure S1, which were generated using the web-enabled tool "PV Analyzer".[72] As expected, PBDTTT-EFT:PC$_{71}$BM ($5.8 \times 10^{-12}$ A) shows a lower reverse bias saturation current than PBDTTT-EFT:Ph2a ($3.6 \times 10^{-10}$ A). Additionally, the PBDTTT-EFT:PC$_{71}$BM device has $n = 1.8$ and $R_s$ = 1.1 $\Omega$ cm$^2$ in contrast to the PBDTTT-EFT:Ph2a device with $n = 2.3$ and $R_s = 1.3$ $\Omega$ cm$^2$. A higher ideality factor and a higher series resistance for the non-fullerene acceptor device indicate that higher recombination currents are present in the Ph2a devices and also likely accounts for the observed "kink" in the *J-V* curve at ~0.80 V.[73]

Figure 1d shows the illuminated *J-V* characteristics for the PBDTTT-EFT:PC$_{71}$BM and PBDTTT-EFT:Ph2a devices that achieved power conversion efficiencies of 8.3% and 4.9%, respectively. The performance of the PBDTTT-EFT:PC$_{71}$BM cell agrees well with literature precedent.[6,67,74,75] Further optimization of the PBDTTT-EFT:Ph2a system in this work led to an increase in OPV performance over previously reported results.[31] A sample of eight PBDTTT-EFT:PC$_{71}$BM devices attained a power conversion efficiency of 8.1 ± 0.2%, open-circuit voltage of 0.79 ± 0.004 V, short-circuit current density of 15.2 ± 0.2 mA cm$^{-2}$, and fill factor of 0.68 ± 0.01. Eight PBDTTT-EFT:Ph2a devices attained a power conversion efficiency of 4.8 ± 0.1%, open-circuit voltage of 0.91 ± 0.003 V, short-circuit current density of 11.0 ± 0.1 mA cm$^{-2}$, and fill factor of 0.48 ± 0.01. Plots of the external quantum efficiency (EQE) spectra for the PBDTTT-EFT:PC$_{71}$BM and PBDTTT-EFT:Ph2a devices and the unnormalized optical absorption spectra are shown in Figure S2a and S2b, which reveals that the higher short-circuit current in PBDTTT-EFT:PC$_{71}$BM derives from a broadband enhancement in EQE over the PBDTTT-EFT:Ph2a system. Indeed, the lower short-circuit currents and fill factors in the non-fullerene acceptor devices suggest that charge recombination is limiting performance in these materials.[19,30,31] Nevertheless, the performance attained here compares well with standard high-performance OPV systems such as PTB7:PC$_{71}$BM.[66,76–79] Thus, these samples are appropriate for further characterization with LFN and impedance spectroscopy.

## 2.2. Correlated Low-Frequency Noise and Impedance Spectroscopy in the Dark

Previously, LFN spectroscopy has been employed to assess the quality of charge percolation networks in organic polymer blends.[55] This technique has also been used to reveal the role of charge traps in organic field-effect transistors[80–84] and organic light-emitting diodes.[85] However, there have only been a handful of LFN studies to date on OPVs. Most of these focused on studying the role of the donor-to-acceptor ratio and annealing conditions on charge transport. Furthermore, these LFN reports have been limited to the canonical poly(3-hexylthiophene-2,5-diyl):[6,6]-phenyl-C$_{61}$-butyric acid methyl ester (P3HT:PC$_{61}$BM) system.[86–89] While there has been an LFN study on (poly[2,6-(4,4-bis-(2-ethylhexyl)-4H-cyclopenta[2,1-b;3,4-b0]-dithiophene)-alt-4,7-(2,1,3-benzothiadiazole)] (PCPDTBT):PC$_{71}$BM OPVs that showed power conversion efficiencies in the range of $1 - 3\%$,[90] LFN has not yet been performed on more recent, higher-performance OPVs such as PTB7:PC$_{71}$BM or devices utilizing emerging non-fullerene acceptors. Since recombination is a known challenge in PDI acceptors,[31] LFN analysis is expected to yield important physical insights in these OPV materials since $1/f$ noise is caused by resistance fluctuations in metals and semiconductors and originates from either mobility fluctuations (Hooge model)[91] or carrier number fluctuations (McWhorter model).[92] This $1/f$ excess noise dominates other sources of noise, such as generation-recombination noise, shot noise, and Johnson-Nyquist (i.e., thermal) noise, at low frequencies. Both Johnson-Nyquist and shot noise show no spectral dependence (characteristic of white noise).[69] Generation-recombination noise is usually characterized by either a single or a few Lorentzians producing sharp spectral transitions or spectral flattening at a characteristic frequency.[93,94] Deviations from strictly $1/f$ behavior (such as $1/f^{1.5}$ or $1/f^{2}$) can be explained by invoking modified models to specific materials.[95,96] Hooge's empirical relation (eq. 2) has been widely invoked to explain noise behavior in a wide variety of materials and devices,

$$S_I = \left(\frac{a_H}{N}\right)\left(\frac{I^{g}}{f^{b}}\right) \tag{2}$$

where $S_I$ is the current power spectral density, $\alpha_H$ the empirical Hooge parameter, $N$ the carrier density, $I$ the mean device current, and $f$ is the frequency. The exponents $\beta$ and $\gamma$ are ideally expected to be equal to 1 and 2, respectively. The Hooge mobility fluctuation model has been previously used to model the noise spectral density in P3HT:PC$_{61}$BM OPVs.[87,88] Building from Hooge's empirical relation,[97] Kleinpenning derived a model to explain V-shaped noise characteristics in silicon $p$-$n$ junction diodes in the dark and under illumination.[98,99] In particular, eq. 3 describes $1/f$ noise in the dark for a diffusion-current dominated $p$-$n$ junction diode (an analogous case to OPVs),

$$S_I(f) = \frac{a_H q I_0}{4 f t}(e^{\frac{qV}{k_B T}} - 1 - \frac{qV}{k_b T}) \tag{3}$$

where $S_I(f)$ is the current power spectral density, $\alpha_H$ is the Hooge parameter, $q$ is the elementary unit of charge, $I_0$ is the reverse bias saturation current, $f$ is frequency, $\tau$ is the charge carrier lifetime, $V$ is the applied bias, $k_B$ is Boltzmann's constant, and $T$ is temperature. While the Kleinpenning $p$-$n$ junction noise model has been successful in describing classical systems such as silicon solar cells, it has not yet been attempted for OPVs.[100–102]

We first extract $\tau$ from impedance spectroscopy to more accurately fit the LFN data to the Kleinpenning model. Impedance spectroscopy measurements were conducted in the dark to obtain the charge carrier lifetimes. Figure 2a and 2b show Cole-Cole plots (–reactance, Z" versus resistance, Z') of the impedance response of a PBDTTT-EFT:PC$_{71}$BM device and a PBDTTT-EFT:Ph2a device, respectively (see Experimental Section for details). In the present study, we employ a proposed equivalent circuit model based on previous work with BHJ P3HT:PCBM OPVs that models both OPV systems well here. In particular, the model assumes the superposition of two partial semicircular arcs to describe the impedance response,[60] as shown in Figure 2c. It incorporates a resistor R$_s$ (series resistance) in series with a resistor R$_{bulk}$ (bulk resistance) and capacitor C$_{geo}$ (geometrical capacitance) in parallel. The R$_{bulk}$||C$_{geo}$ element is also in parallel with a resistor R$_{rec}$ (recombination resistance) and constant phase element (CPE).[103] The R$_s$ element is the device series resistance, which for both samples is generally on the order of ~15-20 $\Omega$, or ~0.9-1.2 $\Omega$ cm$^2$ (the device area is ~ 0.06 cm$^2$), in good agreement with the Shockley diode model fits of the dark $J$-$V$ characteristics. Additionally, R$_{bulk}$||C$_{geo}$ is associated with a bulk resistance and geometrical capacitance (i.e., the electrodes can be conceptualized as parallel plates and the active layer as the dielectric). Lastly, R$_{rec}$||CPE is physically understood to be related to charge transfer events at the donor-acceptor interface, where R$_{rec}$ is the recombination resistance and CPE is related to the chemical capacitance. Note that a constant phase element can be conceptualized as a capacitor with a distribution of relaxation times, which not only simplifies the equivalent circuit model for OPV systems but also has successfully modeled the non-ideal capacitive nature of heterogeneous interfaces in BHJ OPV devices.[103] Eq. 4 shows the chemical capacitance that can be derived from the CPE (eq. 4),

$$C_m = \frac{t_{avg}}{R_{rec}} = \frac{(R_{rec}Q)^{1/m}}{R_{rec}} \qquad (4)$$

where $C_\mu$ is the chemical capacitance, $\tau_{avg}$ is the average of the distribution of relaxation times, $R_{rec}$ is the recombination resistance, $Q$ is the CPE magnitude, and $m$ is an ideality factor that is characteristic of the relaxation time distribution. An ideal capacitor has $m = 1$, with $m$ ranging from 0.54 to 0.75 in this work.

As excess charge is generated and stored in the active layer (e.g., under an applied electrical or optical bias), the chemical capacitance increases and the recombination resistance decreases. By definition, the geometrical capacitance is expected to be nearly constant with applied bias. In contrast, the chemical capacitance is expected to increase exponentially with increasing bias as excess electrons populate the electron acceptor LUMO. These trends help distinguish between the two $R$-$C$ elements when examining bias-dependent impedance data. The characteristic charge-discharging time constant of each $R$-$C$ element is obtained by multiplying the values together. The greater time constant is associated with the charge carrier lifetime, since the larger time constant limits charge extraction.

Plots of $R_{bulk}$ and $C_{geo}$ are shown in Figure 3a for PBDTTT-EFT:PC$_{71}$BM and Figure 3b for PBDTTT-EFT:Ph2a. The increase in $C_{geo}$ at larger biases has been previously observed, and is likely related to a modified relative permittivity due to a large injected carrier population.[60] The lower bias $C_{geo}$ values and film thicknesses (measured by profilometry) can be used to calculate the dielectric constant, $\varepsilon_r$, assuming a parallel-plate model. For PBDTTT-EFT:PC$_{71}$BM, the 2.72 nF geometrical capacitance and 98 nm film thickness yield $\varepsilon_r = 5.0$. For PBDTTT-EFT:Ph2a, with a geometrical capacitance of 2.10 nF and film thickness of 70 nm, we obtain $\varepsilon_r = 2.8$. These values are comparable to previous reports,[60] and well within the range of theoretical capacitance values possible for organic materials.[104]

Figure 3c and 3d show the extracted recombination resistance and chemical capacitance for PBDTTT-EFT:PC$_{71}$BM and PBDTTT-EFT:Ph2a, respectively. Overall, the PBDTTT-EFT:PC$_{71}$BM results are similar to the P3HT:PC$_{61}$BM system,[60,65,105] with $R_{rec}$ decreasing monotonically with increasing bias. The chemical capacitance $C_\mu$ exhibits a power law dependence on sample bias with a scaling exponent of ~14. While $C_\mu$ is expected to increase exponentially, slight deviations from an exponential dependence have been observed previously and attributed to electron trapping in the conduction band.[106–109] While $R_{rec}$ decreases monotonically with increasing bias for PBDTTT-EFT:Ph2a as expected, the chemical capacitance exhibits greater invariance at lower biases and decreases significantly from 0.60 V to 0.80 V. Note that decreasing and/or negative chemical capacitance at increasing forward bias, have been observed previously in organic solar cells and attributed to a decrease in the series resistance under larger forward bias that modulates the conductivity, such that additional carrier injection leads to more discharging as opposed to charge accumulation.[110] Since the average time constant associated with $R_{bulk}\|C_{geo}$ (~400 µs at low forward bias) is greater than the time constant for $R_{rec}\|CPE$ (~100 µs at low forward bias) for PBDTTT-EFT:PC$_{71}$BM, $R_{bulk}\|C_{geo}$ is the limiting element to charge collection and is therefore assigned to be the charge carrier lifetime, $\tau$. In contrast, for PBDTTT-EFT:Ph2a, the average time constant associated with $R_{bulk}\|C_{geo}$ (~200 µs at low forward bias) is lower than the time constant for $R_{rec}\|CPE$ (~800 µs at low forward bias), and thus $R_{rec}\|CPE$ is assigned to be the charge carrier lifetime. In the dark, $R_{rec}\|CPE$ is likely associated with resistance to interfacial charge transfer and interfacial charge accumulation.[106,111] Since computational studies have argued that there is more efficient interfacial charge transfer between fullerenes than in PDIs,[40,44] it is not surprising to see that $R_{rec}\|CPE$ plays a more significant role in the PBDTTT-EFT:Ph2a system. The charge carrier lifetime, $\tau$, is shown in Figure 3e and 3f for PBDTTT-EFT:PC$_{71}$BM and PBDTTT-EFT:Ph2a, respectively. Note that the lifetimes for both samples are within the expected range (i.e., hundreds of µsec),[40] and that the values for PBDTTT-EFT:Ph2a are approximately the same as the PBDTTT-EFT:PC$_{71}$BM lifetimes. As shown in the carrier lifetime plots, an exponential decay function provides a good fit to both sets of lifetime data in the dark.

After extracting the charge carrier lifetimes, LFN measurements were next performed on these OPV systems. Figure 4a and 4b show current noise spectral density ($S_I$) as a function of frequency at biases of -1 V and +0.90 V for a PBDTTT-EFT:PC$_{71}$BM device and a PBDTTT-EFT:Ph2a device in the dark, respectively. Under reverse bias, $S_I$ is significantly lower (around a factor of ~100x across the frequency bandwidth measured) for PBDTTT-EFT:PC$_{71}$BM versus PBDTTT-EFT:Ph2a. While differences between these systems may be attributable to injection related noise,[112,113] we note that the higher $S_I$ in PBDTTT-EFT:Ph2a correlates with its lower power conversion efficiency, higher dark diode current, larger diode ideality factor, and lower recombination resistance (at equivalent biases). This observation indicates that Ph2a-based BHJ OPVs have a greater proportion of defects in comparison to PC$_{71}$BM-based BHJ OPVs. Similar trends have been observed in P3HT:PC$_{61}$BM OPVs, where the dark diode current and $S_I$ concurrently decrease when the active layer is annealed at higher temperatures, reducing film defect density.[58] Moreover, lower power conversion efficiency and noise amplitude have also been observed in silicon solar cells irradiated by proton irradiation.[101] Furthermore, a lower magnitude of $S_I$ has been also correlated with higher power conversion efficiency and lower defect density in silicon solar cells.[114] This validates the higher performance and lower noise magnitude observed in PBDTTT-EFT–PC$_{71}$BM. These trends can be further verified by calculating the specific detectivity, which is indicative of the photodetector quality of these devices. The detectivity is calculated using eq. 5,[115]

$$D* = \frac{\lambda e \sqrt{A} \times EQE}{hc\sqrt{S_I}} \quad (5)$$

where $D*$ is the detectivity, $\lambda$ the wavelength, $A$ the device area, $EQE$ the external quantum efficiency, $h$ Planck's constant, $c$ the speed of light, and $S_I$ is the current power spectral density. More accurate values of $S_I$ are obtained from LFN measurements in contrast to inferring the value from the measured dark diode current, which only accounts for shot noise and neglects $1/f$ noise.[50–53] Values for $S_I$ are chosen at the same frequency as employed for the EQE measurements, 30 Hz. Using $\sqrt{S_I}$ values of $3.14 \times 10^{-11}$ A $Hz^{-1/2}$ for PBDTTT-EFT:PC$_{71}$BM and $6.65 \times 10^{-10}$ A $Hz^{-1/2}$ for PBDTTT-EFT–Ph2a, noting the device area of ~0.06 cm$^2$, and choosing $\lambda = 700$ nm (with EQE values of 0.70 and 0.55 for PBDTTT-EFT:PC$_{71}$BM and PBDTTT-EFT–Ph2a, respectively), the calculated detectivity values are $3.08 \times 10^9$ $Hz^{-1/2}$ cm $W^{-1}$ for PBDTTT-EFT:PC$_{71}$BM and $1.05 \times 10^8$ $Hz^{-1/2}$ cm $W^{-1}$ for PBDTTT-EFT–Ph2a. Evidently, lower detectivity values are correlated with higher defect densities.[54] While the PBDTTT-EFT:Ph2a detectivity is lower, note that there is interest in developing organic photodetectors with red-shifted absorption in comparison to fullerene-based photodetectors, for near-infrared detection.[54] Since PDIs are more easily synthetically tuned than their fullerene counterparts,[30] the photodetectivity values observed here indicate the potential for near-infrared active PDI-based organic photodetectors.

Further analysis of the LFN data shows that both samples in Figure 4a,b exhibit $1/f^\beta$ behavior under reverse bias, where $\beta$ is 1.02 and 1.36 for PBDTTT-EFT:PC$_{71}$BM and PBDTTT-EFT:Ph2a, respectively. For the entire sample set, three PBDTTT-EFT:PC$_{71}$BM devices exhibited $\beta = 1.01 \pm 0.03$ at -1 V and three PBDTTT-EFT:Ph2a devices exhibited $\beta = 1.19 \pm 0.14$ at -1 V. The forward bias spectra exhibit $1/f^\beta$ behavior for $f < 10$ Hz, with $\beta = 1.21$ for PBDTTT-EFT:PC$_{71}$BM ($\beta = 1.13 \pm 0.11$ for three devices) and $\beta = 1.25$ for PBDTTT-EFT:Ph2a ($\beta = 1.19 \pm 0.06$ for three devices). Previously, conductive P3HT films mixed with insulating polystyrene have shown increased charge trapping, leading to higher values of $\beta$.[55] Correspondingly, the higher $\beta$ values observed in PBDTTT-EFT:Ph2a OPVs suggest that less efficient charge transport and more recombination are present in this system relative to PBDTTT-EFT:PC$_{71}$BM. Methods to potentially mitigate these recombination events and tune OPV film morphology include optimized processing via thermal or solvent annealing[116–118] or the incorporation of processing additives,[119,120] which could potentially enhance the performance of the PBDTTT-EFT:Ph2a system. Furthermore, tuning the molecular weight of the donor polymer has been shown via transmission electron microscopy (TEM) to significantly impact donor-acceptor blend aggregation, with a high degree of donor-acceptor interfacial overlap minimizing recombination and optimizing device performance.[121] Therefore, the observed correlation between $\beta$ values extracted from low-frequency noise measurements and interfacial overlap observed via TEM suggests that low-frequency noise measurements have the potential to reveal performance-limiting structural mechanisms that other destructive techniques, such as TEM, can also provide.

While the reverse-biased LFN spectra in Figure 4a and 4b exhibit typical $1/f$ behavior, the forward bias LFN spectra flatten at frequencies exceeding 10 Hz. This flattening in the LFN spectra was also observed by Landi in P3HT:PC$_{61}$BM OPVs,[88] albeit at higher frequencies than what is observed here. Note that additional annealing of the aforementioned P3HT:PC$_{61}$BM OPVs upshifted the corner flattening frequency beyond the experimental bandwidth. Since the devices in the present study are not annealed during fabrication and did not undergo any post-fabrication annealing, it is not surprising that the corner flattening frequency is downshifted relative to P3HT:PC$_{61}$BM OPVs. One possible culprit for the flattening is shot noise, which can dominate low-frequency noise at high driving currents. Shot noise $S_I$ scales as $\sim I^1$.[69] Plots of Log($S_I$) vs Log($I$) for PBDTTT-EFT:PC$_{71}$BM and PBDTTT-EFT:Ph2a at high forward biases (see Figure S3a and S3b) reveal the current exponent $\gamma$ for PBDTTT-EFT:PC$_{71}$BM to be $1.66 \pm 0.10$ ($1.67 \pm 0.11$ for three devices) and $\gamma$ for PBDTTT-EFT:Ph2a to be $1.59 \pm 0.09$ ($1.54 \pm 0.12$ for three devices). These values rule out shot noise alone as the origin of the spectral flattening. Another related possibility is Johnson-Nyquist noise.[88] The magnitude of Johnson-Nyquist noise can be calculated using eq. 6[122]

$$S_I = \frac{4k_B T}{R} \qquad (6)$$

where $S_I$ is the noise spectral density, $k_B$ is Boltzmann's constant, $T$ is temperature, and $R$ is the device resistance. Local linear fits of the device current-voltage plots were used to obtain values for $R$ at the same high forward biases used for noise measurements. In particular, the $R$ values obtained for PBDTTT-EFT:PC$_{71}$BM and PBDTTT-EFT:Ph2a were 25 Ω and 250 Ω respectively, leading to an expected Johnson-Nyquist noise level of ~$10^{-22}$ - $10^{-23}$ A$^2$/Hz. However, this value is significantly lower than the measured noise levels (~$10^{-15}$ A$^2$/Hz) under high forward bias. Therefore, thermal noise also does not account for the observed spectral features. Another possible source of this flattening may be carrier generation-recombination noise that shows flattening of the noise spectral density before a sharp downward turn (Lorentzian) at a corner frequency beyond the bandwidth explored here.[93] Generation-recombination noise has been observed before in similar amorphous systems that possess a high degree of structural disorder (e.g., amorphous silicon solar cells[123] and pentacene thin-film transistors[94]). In this report, we do not pursue the source of this noise at higher frequencies further and instead focus the analysis in the low-frequency regime where $1/f$ behavior is firmly established.

We next discuss the dependence of LFN on voltage bias to further elucidate its origin in these devices. Figure 4c and 4d show plots of $S_I$ as a function of voltage at $f = 2$ Hz for the PBDTTT-EFT:PC$_{71}$BM and PBDTTT-EFT:Ph2a systems, in the dark, respectively. In both cases, the voltage dependence of the noise is a V-shaped curve (in the log-linear plot) with a vertex at 0.0 V. The plot for PBDTTT-EFT:Ph2a is visibly more symmetric than for PBDTTT-EFT:PC$_{71}$BM, which is related to the fact that dark current density-voltage data are also more symmetric for PBDTTT-EFT:Ph2a (Figure 1c). Additionally, as shown in Figure S4, $S_I$ does not show an I$^2$ dependence, with a current exponent of 0.99 ± 0.09 for PBDTTT-EFT:PC$_{71}$BM and a current exponent of 1.09 ± 0.05 for PBDTTT-EFT:Ph2a, suggesting the noise does not simply scale with device current. Previously, V-shaped voltage dependence of noise was observed in silicon $p$-$n$ homojunction photodetectors and solar cells and described by Kleinpenning's diffusion-current dominated *p-n* junction diode model (eq. 3).[69,98,124,125] In these devices, the Kleinpenning model has been fit with forward and reverse bias diode data using a constant value for $\tau$.[69] However, as shown in Figure 3e and 3f, the charge carrier lifetimes in OPV systems vary as a function of applied positive bias. Therefore, we have modified the Kleinpenning model to fit the bias-dependent $S_I$ data by taking a negative exponential functional form for $\tau$ as deduced from impedance spectroscopy. This approach yields fits to the PBDTTT-EFT:PC$_{71}$BM and PBDTTT-EFT:Ph2a noise spectral density–voltage data (Figure 4c and 4d). The PBDTTT-EFT:Ph2a system exhibits a higher diode ideality factor and series resistance than PBDTTT-EFT:PC$_{71}$BM, which likely accounts for the better fit for PBDTTT-EFT:PC$_{71}$BM. Previously, the metric $\alpha_H/\tau$ has been reported to range from $10^5$-$10^8$ s$^{-1}$ for P3HT:PC$_{61}$BM OPVs.[86] Using extracted fit coefficients, the metric $\alpha_H/\tau$ is ~3 × $10^7$ s$^{-1}$ for PBDTTT-EFT:PC$_{71}$BM and ~3 × $10^6$ s$^{-1}$ for PBDTTT-EFT:Ph2a, which is in reasonable agreement with the values reported for P3HT:PC$_{61}$BM. Thus, the Kleinpenning model, originally developed for crystalline and homogenous *p-n* junctions, also describes the noise characteristics of highly disordered and heterogeneous BHJ OPV systems, utilizing conventional fullerene and emerging non-fullerene acceptors.

## 2.3. Correlated Low-Frequency Noise and Impedance Spectroscopy under Illumination

While correlated LFN measurements and impedance spectroscopy of OPVs in the dark provide useful insights into charge dynamics, in-situ measurements of LFN and impedance under illumination have the potential to further elucidate the performance-limiting mechanisms of these OPV devices. While only one report has studied 1/*f* noise of OPVs under illumination,[90] impedance spectroscopy has been more widely applied to OPVs under operating conditions, including the canonical materials systems P3HT:PC$_{61}$BM[60,64,65] and PTB7:PC$_{71}$BM.[126] Here, we report correlated LFN and impedance analysis of both PBDTTT-EFT:PC$_{71}$BM and PBDTTT-EFT:Ph2a OPV samples under AM1.5G 1 sun illumination.

Figure 5a and 5b show Cole-Cole plots for a PBDTTT-EFT:PC$_{71}$BM device and a PBDTTT-EFT:Ph2a device under 1 sun, respectively. The reactance and the resistance are reduced by a factor of ~100x when photo-generated carriers are present. While both the bulk and recombination resistance decrease under illumination, the bulk resistance drastically reduces. Additionally, the chemical capacitance values under illumination exhibit a power law dependence as a function of applied bias. If the geometrical capacitance was a dominant factor here, the capacitance should be invariant as a function of applied bias. Therefore, the contribution of this *R-C* element to the overall impedance response is effectively negligible. As a result, impedance data under illumination were fit to a simpler equivalent circuit model in comparison to the model used for dark measurements (Figure 5c).[126] Specifically, this model incorporates a resistor ($R_s$) in series with a resistor ($R_{rec}$ – recombination resistance) and capacitor ($C_\mu$ - chemical capacitance) in parallel. The bias dependence of R$_{rec}$ and C$_\mu$ is shown in Figure 6a for PBDTTT-EFT:PC$_{71}$BM and Figure 6b for PBDTTT-EFT:Ph2a. As sample bias increases, R$_{rec}$ decreases monotonically, which has been previously observed in P3HT:PC$_{61}$BM OPVs.[60] C$_\mu$ exhibits a power law dependence for both systems, where the scaling exponent is ~2.6 for PBDTTT-EFT:PC$_{71}$BM and ~1.5 for PBDTTT-EFT:Ph2a. While C$_\mu$ is expected to show exponential behavior, the deviation observed here is the possible result of electron trapping, as has been previously noted.[106–109] Notably, the significantly lower values for R$_{rec}$ under illumination (by a factor of ~10x) in comparison to the dark data imply a decreased lifetime due to the large population of photo-generated carriers. Indeed, $\tau$ for both PBDTTT-EFT:PC$_{71}$BM (Figure 6c) and PBDTTT-EFT:Ph2a OPVs (Figure 6d) is reduced by a factor of ~10x under illumination. Additionally, the $\tau$ values for PBDTTT-EFT:PC$_{71}$BM are greater than for PBDTTT-EFT:Ph2a at low bias. Larger lifetimes correlate with higher power conversion efficiency in P3HT:PC$_{61}$BM OPVs,[60] a trend also observed here for the illuminated data.

Lastly, we discuss LFN measurements under 1 sun illumination. Figure 7a and 7b show noise spectral density as a function of frequency at -1 V and +1 V, for PBDTTT-EFT:PC$_{71}$BM and PBDTTT-EFT:Ph2a devices, respectively. Qualitatively, both dark and illuminated data reveal similar noise spectral behavior under reverse bias (Figure 4a, 4b, 7a, 7b). Under illumination, both sets of devices exhibit $1/f^{\beta}$ behavior with $\beta = 1.05$ for PBDTTT-EFT:PC$_{71}$BM ($\beta = 1.17 \pm 0.13$ for three devices) and $\beta = 1.08$ for PBDTTT-EFT:Ph2a ($\beta = 1.17 \pm 0.15$ for three devices). The forward bias spectra exhibit $1/f^{\beta}$ behavior for $f < 10$ Hz, with $\beta = 1.50$ for PBDTTT-EFT:PC$_{71}$BM ($\beta = 1.82 \pm 0.23$ for three devices) and $\beta = 1.47$ for PBDTTT-EFT:Ph2a ($\beta = 1.38 \pm 0.23$ for three devices). Figure S5a and S5b show Log($S_I$) vs Log($I$) under high forward bias for PBDTTT-EFT:PC$_{71}$BM and PBDTTT-EFT–Ph2a, respectively. The current exponent for PBDTTT-EFT:PC$_{71}$BM is $\gamma = 0.71 \pm 0.07$ ($\gamma = 1.66 \pm 1.4$ for three devices) and for PBDTTT-EFT:Ph2a is $\gamma = 2.37 \pm 0.08$ ($\gamma = 1.59 \pm 0.64$ for three devices). Note that the magnitude of $S_I$ increases for each OPV system under illumination, which likely reflects the increased carrier density in comparison to the dark condition. The presence of more carriers is expected to fill trap states in the OPV BHJ film while also increasing scattering,[127–129] raising the overall noise level. A similar trend was observed previously in PCPDTBT:PC$_{71}$BM OPVs.[90] An increase in $S_I$ is also observed in silicon photodiodes under illumination.[69] It is especially likely for noise to increase under illumination in OPVs given the high degree of structural disorder present in these devices.[130] In contrast to the LFN behavior in the dark, the reverse bias values of $\beta$ under illumination become comparable for both of the present OPV systems. This behavior indicates that both systems give rise to a more similar densities-of-states under illumination.[55] Furthermore, the illuminated forward bias noise spectral density versus frequency spectra show similar behavior to the dark spectra, particularly flattening at frequencies exceeding 10 Hz, which cannot be assigned to either thermal or shot noise. Therefore, similar to the dark data, this flattening may be similarly arising from a generation-recombination noise source.

$S_I$ under illumination is plotted as a function of voltage in Figure 7c for PBDTTT-EFT:PC$_{71}$BM and in Figure 7d for PBDTTT-EFT–Ph2a. In sharp contrast to the dark data, the light $S_I$ versus voltage data are nearly independent of applied bias. $S_I$ for PBDTTT-EFT:Ph2a is on the order of ~$10^{-14}$ A$^2$/Hz and for PBDTTT-EFT:PC$_{71}$BM $S_I$ is on the order of ~$10^{-15}$ A$^2$/Hz over the bias range measured. Note that the higher $S_I$ values for PBDTTT-EFT:Ph2a correlate with its lower PCE. The invariance in $S_I$ as a function of voltage suggests that the noise induced by the photocurrent dominates the noise spectral density. Under illumination, the device current values are more constant over the bias window measured, which could account for the noise invariance at varying biases. Moreover, the current magnitudes in PBDTTT-EFT:PC$_{71}$BM and PBDTTT-EFT:Ph2a are more comparable under illumination, which partly explains similar values of $S_I$ in contrast to the dark $S_I$ data. The measured $S_I$ converges to similar values in the dark and under illumination at higher forward biases, where the current magnitudes become comparable. The recombination resistance, chemical capacitance, and charge carrier lifetimes extracted from impedance measurements were also more comparable under illumination than in the dark. Therefore, both LFN and impedance measurements under illumination show that the presence of photo-generated carriers governs the spectral response. This observation is not surprising since illumination dominates the current-voltage response of a solar cell, particularly below the turn-on voltage.

## 3. Conclusions

In this work, BHJ OPV device performance, charge transport, and current fluctuations have been quantified in devices utilizing a conventional fullerene acceptor and an emerging dimeric PDI acceptor. Standard current-voltage characterization reveals that PDI-based OPV performance is somewhat lower than devices using conventional fullerenes. These results are then correlated with low-frequency noise measurements in the dark, which reveal a narrower density-of-states profile in the non-fullerene acceptor OPVs. This result implies that a greater degree of recombination, and thus disorder, is present in the non-fullerene OPVs. Furthermore, the elucidation of charge carrier lifetimes from impedance measurements enable the dark noise data to be fit to the Kleinpenning model for the first time. Measurements under illumination reveal increased noise spectral density relative to the dark, by approximately four orders of magnitude for fullerene-based OPVs and three orders of magnitude for PDI-based OPVs. Furthermore, an inverse correlation is found between noise spectral density and solar cell efficiency. Overall, this work establishes that in-situ correlated LFN and impedance spectroscopy are powerful analytical tools that can offer physical insight into OPV function beyond standard current-voltage characterization, and have the potential to serve as cost-effective, non-destructive methods to correlate structure-performance relationships, which could guide the optimization of emerging photovoltaic materials.

## 4. Experimental Section

### 4.1. Substrate Cleaning

Patterned indium tin oxide (ITO, 20 $\Omega$ sq$^{-1}$) on glass was procured from Thin Film Devices, Inc. The substrates were sonicated at 50 ℃ in aqueous detergent for 30 min, and then for 20 min in deionized water, methanol, isopropanol, and acetone baths in succession. If stored (in ambient) prior to use, the ITO substrates were sonicated at 50 ℃ for 10 min successively in methanol, isopropanol, and acetone immediately preceding device fabrication. Finally, the ITO substrates were exposed to a 5 min air plasma treatment at 18 W under rough vacuum (~300 mTorr).

### 4.2. Zinc Oxide Film Deposition

A ~0.50 M zinc oxide (ZnO) sol gel solution was produced by combining 220 mg of zinc acetate dihydrate (Zn(CH$_3$COO)$_2$•2H$_2$O), 99.999%, Sigma-Aldrich), 62 mg of ethanolamine (NH$_2$CH$_2$CH$_2$OH, ≥99%, Sigma-Aldrich), and 2 mL of 2-methoxyethanol (CH$_3$O-CH$_2$CH$_2$OH, anhydrous, 99.8%, Sigma-Aldrich).[71] This solution was sonicated for approximately 5 min at ambient temperature until the solids were visibly dissolved and aged overnight. Then, the ZnO sol gel solution was filtered through a 0.45 µm polyvinylidene difluoride (PVDF) filter and spun-cast at 7000 RPM for 30 sec. Following spin-casting, substrates were promptly placed on a hot plate at 170 ℃ and annealed for 10 min. The ZnO substrates were immediately transferred to an argon-filled glovebox after annealing.

*4.3. Active Layer Deposition*

A poly[4,8-bis(5-(2-ethylhexyl)thiophen-2-yl)benzo[1,2-b;4,5-b′]dithiophene-2,6-diyl-alt-(4-(2-ethylhexyl)-3-fluorothieno[3,4-b]thiophene-)-2-carboxylate-2-6-diyl)] (PBDTTT-EFT, 1-Material)–[6,6]-phenyl-$C_{71}$-butryic acid methyl ester ($PC_{71}BM$, American Dye Source, Inc.) solution was prepared by dissolving PBDTTT-EFT (13 mg/mL) and $PC_{71}BM$ (26 mg/mL) powder in a solution of dichlorobenzene:1,6-di-iodohexane (97.5:2.5 vol%) with a total loading of 39 mg/mL. This solution was stirred overnight on a hot plate at 75 ℃ in an argon-filled glovebox prior to use. Next, 20 µL of the PBDTTT-EFT:$PC_{71}BM$ solution was dispensed onto a ZnO substrate, which was then spun cast at 1700 RPM for 20 sec. These samples were placed in covered Petri dishes for approximately 1.5 hours.

A PBDTTT-EFT–[1,2:3,4]-bis-[*N,N*′-bis-perylenediimide-1,12-yl]-benzene (Ph2a) solution was prepared by dissolving PBDTTT-EFT (4 mg/mL) and Ph2a (6 mg/mL) in a solution of chloroform:diphenyl ether (99:1 vol%) with a total loading of 10 mg/mL. The synthesis of Ph2a is described elsewhere.[31] This solution was stirred overnight on a hot plate at 75 ℃ in an argon-filled glovebox prior to use. Next, the PBDTTT-EFT:Ph2a active layer films were prepared by dispensing 20 µL of solution onto an already-spinning substrate (1500 RPM, 30 sec). Film thicknesses were measured with profilometry (Dektak 150 Stylus Surface Profiler). Reported film thicknesses are an average of five measurements.

*4.4. Contact Deposition*

Portions of the substrate were etched with a dry cotton swab following active layer spin casting to define an area for a cathode bus bar on each substrate. Subsequently, the substrates were placed in a glovebox-enclosed thermal evaporator. A shadow mask was placed on the substrates to define a cathode bus bar and anode contact such that there were four 0.06 cm$^2$ devices per substrate. Next, the chamber was pumped down to ~5 x 10$^{-6}$ Torr. Then, MoO$_3$ (Puratronic, 99.9995%) and silver (W. E. Mowrey, 99.99%) were iteratively deposited without breaking vacuum at respective deposition rates of ~0.2 Å s$^{-1}$ and 1.5-2 Å s$^{-1}$. After removal from the vacuum chamber, samples were encapsulated in a UV chamber (Electro-Lite) using glass slides and UV-curable ELC-2500 epoxy prior to removal from the glovebox.

*4.5. Solar Cell Evaluation*

A solar cell analyzer (Class A Spectra-Nova Technologies), equipped with a xenon arc lamp and AM1.5G filter, was used to collect current-voltage data in the dark and at 1 sun illumination. Calibration with a monocrystalline silicon diode fitted with a KG3 filter brought spectral mismatch close to unity. A Newport Oriel® Quantum Efficiency Measurement Kit recorded external quantum efficiency (EQE) data. The EQE data were integrated against the solar spectrum using the Open Photovoltaics Analysis Platform program,[131] which was used to correct the short-circuit current density values measured by the solar simulator (additionally the EQE-derived short-circuit current density values were within <10% of the values measured by the solar simulator). Dark diode data were also collected and analyzed within the Shockley *p-n* junction diode model using the web-enabled tool "PV Analyzer."[72]

*4.6. Impedance Spectroscopy*

Impedance spectroscopy measurements were conducted using a Solartron Model 1260A impedance/gain phase analyzer. The sample impedance was measured over a frequency range of 1 Hz to 1 MHz, and the oscillation amplitude did not exceed 50 mV. Impedance spectra were recorded in the dark and under AM1.5G 1 sun illumination using a solar simulator (Newport).

*4.7. Low-Frequency Noise Measurements*

Low-frequency noise measurements were conducted using a low noise current pre-amplifier (1212 DL Instruments) and a spectrum analyzer (Stanford Research Systems, SR760). A Keithley 2400 source-measurement unit was employed to bias devices during noise measurements and to concurrently measure device current.[93,132] Comparative noise measurements performed with a DC battery revealed no additional noise from instrumentation in the conditions used in this study. Additionally, a solar simulator (Newport, Inc.) was used to expose the samples to AM1.5G 1 sun illumination to conduct low-frequency noise measurements under simulated sunlight, and to collect current-voltage curves in the dark and light to adjust the sensitivity on the current pre-amplifier accordingly.

**Supporting Information**

Supporting Information is available from the Wiley Online Library or from the author.

**Acknowledgements**

This work was supported as part of the Argonne-Northwestern Solar Energy Research (ANSER) Center, an Energy Frontier Research Center funded by the U. S. Department of Energy, Office of Science, Basic Energy Sciences under Award No. DE-SC0001059. The Institute for Sustainability and Energy at Northwestern (ISEN) provided partial equipment funding. K.A.L. acknowledges a graduate research fellowship from the National Science Foundation. H.N.A. acknowledges support from a NASA Space Technology Research Fellowship (NSTRF, #NNX11AM87H). This work made use of the Keck-II facility of the NUANCE Center at Northwestern University, which has received support from the Soft and Hybrid Nanotechnology Experimental (SHyNE) Resource (NSF NNCI-1542205); the NSF-MRSEC program (NSF DMR-1121262); the International Institute for Nanotechnology (IIN); the Keck Foundation; and the State of Illinois. We also thank N. D. Eastham and Prof. A. S. Dudnik for helpful discussions.

**References**

[1]    C. W. Tang, *Appl. Phys. Lett.* **1986**, *48*, 183.
[2]    M. Graetzel, R. A. J. Janssen, D. B. Mitzi, E. H. Sargent, *Nature* **2012**, *488*, 304.
[3]    K. A. Mazzio, C. K. Luscombe, *Chem. Soc. Rev.* **2014**, *44*, 78.
[4]    G. Yu, J. Gao, J. C. Hummelen, F. Wudl, A. J. Heeger, *Science* **1995**, *270*, 1789.
[5]    S. R. Forrest, *MRS Bull.* **2005**, *30*, 28.
[6]    J.-D. Chen, C. Cui, Y.-Q. Li, L. Zhou, Q.-D. Ou, C. Li, Y. Li, J.-X. Tang, *Adv. Mater.* **2015**, *27*, 1035.

[7]     Y. Liu, J. Zhao, Z. Li, C. Mu, W. Ma, H. Hu, K. Jiang, H. Lin, H. Ade, H. Yan, *Nat. Commun.* **2014**, *5*, 5293.

[8]     D. Konios, C. Petridis, G. Kakavelakis, M. Sygletou, K. Savva, E. Stratakis, E. Kymakis, *Adv. Funct. Mater.* **2015**, *25*, 2213.

[9]     S. Hong, H. Kang, G. Kim, S. Lee, S. Kim, J.-H. Lee, J. Lee, M. Yi, J. Kim, H. Back, J.-R. Kim, K. Lee, *Nat. Commun.* **2016**, *7*, 10279.

[10]    Y. H. Kim, C. Sachse, M. L. Machala, C. May, L. Müller-Meskamp, K. Leo, *Adv. Funct. Mater.* **2011**, *21*, 1076.

[11]    A. Anctil, C. W. Babbitt, R. P. Raffaelle, B. J. Landi, *Environ. Sci. Technol.* **2011**, *45*, 2353.

[12]    R. Po, A. Bernardi, A. Calabrese, C. Carbonera, G. Corso, A. Pellegrino, *Energy Environ. Sci.* **2014**, *7*, 925.

[13]    Y. Lin, F. Zhao, Q. He, L. Huo, Y. Wu, T. C. Parker, W. Ma, Y. Sun, C. Wang, D. Zhu, A. J. Heeger, S. R. Marder, X. Zhan, *J. Am. Chem. Soc.* **2016**, *138*, 4955.

[14]    W. Zhao, D. Qian, S. Zhang, S. Li, O. Inganäs, F. Gao, J. Hou, *Adv. Mater.* **2016**, *28*, 4734.

[15]    Suman, V. Gupta, A. Bagui, S. P. Singh, *Adv. Funct. Mater.* **2017**, *27*, 1603820.

[16]    Y. Lin, J. Wang, Z.-G. Zhang, H. Bai, Y. Li, D. Zhu, X. Zhan, *Adv. Mater.* **2015**, *27*, 1170.

[17]    Y. Lin, Q. He, F. Zhao, L. Huo, J. Mai, X. Lu, C.-J. Su, T. Li, J. Wang, J. Zhu, Y. Sun, C. Wang, X. Zhan, *J. Am. Chem. Soc.* **2016**, *138*, 2973.

[18]    S. Dai, F. Zhao, Q. Zhang, T.-K. Lau, T. Li, K. Liu, Q. Ling, C. Wang, X. Lu, W. You, X. Zhan, *J. Am. Chem. Soc.* **2017**, *139*, 1336.

[19]    Y. Lin, X. Zhan, *Mater. Horiz.* **2014**, *1*, 470.

[20]    X. Zhan, Z. Tan, B. Domercq, Z. An, X. Zhang, S. Barlow, Y. Li, D. Zhu, B. Kippelen, S. R. Marder, *J. Am. Chem. Soc.* **2007**, *129*, 7246.

[21]    Y. Lin, Y. Wang, J. Wang, J. Hou, Y. Li, D. Zhu, X. Zhan, *Adv. Mater.* **2014**, *26*, 5137.

[22]    E. Kozma, M. Catellani, *Dyes Pigments* **2013**, *98*, 160.

[23]    C. Li, H. Wonneberger, *Adv. Mater.* **2012**, *24*, 613.

[24]    R. Schmidt, M. M. Ling, J. H. Oh, M. Winkler, M. Könemann, Z. Bao, F. Würthner, *Adv. Mater.* **2007**, *19*, 3692.

[25]    F. Würthner, M. Stolte, *Chem. Commun.* **2011**, *47*, 5109.

[26]    X. Zhan, A. Facchetti, S. Barlow, T. J. Marks, M. A. Ratner, M. R. Wasielewski, S. R. Marder, *Adv. Mater.* **2011**, *23*, 268.

[27]    J. E. Anthony, *Chem. Mater.* **2011**, *23*, 583.

[28]    Z. Chen, A. Lohr, C. R. Saha-Möller, F. Würthner, *Chem. Soc. Rev.* **2009**, *38*, 564.

[29]    A. Sharenko, C. M. Proctor, T. S. van der Poll, Z. B. Henson, T.-Q. Nguyen, G. C. Bazan, *Adv. Mater.* **2013**, *25*, 4403.

[30]    P. E. Hartnett, A. Timalsina, H. S. S. R. Matte, N. Zhou, X. Guo, W. Zhao, A. Facchetti, R. P. H. Chang, M. C. Hersam, M. R. Wasielewski, T. J. Marks, *J. Am. Chem. Soc.* **2014**, *136*, 16345.

[31]    P. E. Hartnett, H. S. S. R. Matte, N. D. Eastham, N. E. Jackson, Y. Wu, L. X. Chen, M. A. Ratner, R. P. H. Chang, M. C. Hersam, M. R. Wasielewski, T. J. Marks, *Chem. Sci.* **2016**, DOI: 10.1039/C5SC04956C.

[32]    D. Meng, D. Sun, C. Zhong, T. Liu, B. Fan, L. Huo, Y. Li, W. Jiang, H. Choi, T. Kim, J. Y. Kim, Y. Sun, Z. Wang, A. J. Heeger, *J. Am. Chem. Soc.* **2016**, *138*, 375.

[33]    D. Meng, H. Fu, C. Xiao, X. Meng, T. Winands, W. Ma, W. Wei, B. Fan, L. Huo, N. L. Doltsinis, Y. Li, Y. Sun, Z. Wang, *J. Am. Chem. Soc.* **2016**, *138*, 10184.

[34]    Y. Zhong, M. T. Trinh, R. Chen, G. E. Purdum, P. P. Khlyabich, M. Sezen, S. Oh, H. Zhu, B. Fowler, B. Zhang, W. Wang, C.-Y. Nam, M. Y. Sfeir, C. T. Black, M. L. Steigerwald, Y.-L. Loo, F. Ng, X.-Y. Zhu, C. Nuckolls, *Nat. Commun.* **2015**, *6*, 8242.

[35]    R. D. Pensack, C. Guo, K. Vakhshouri, E. D. Gomez, J. B. Asbury, *J. Phys. Chem. C* **2012**, *116*, 4824.

[36]    G. J. Dutton, S. W. Robey, *Phys. Chem. Chem. Phys.* **2015**, *17*, 15953.

[37]    J. M. Szarko, B. S. Rolczynski, S. J. Lou, T. Xu, J. Strzalka, T. J. Marks, L. Yu, L. X. Chen, *Adv. Funct. Mater.* **2014**, *24*, 10.

[38]    H. M. Heitzer, B. M. Savoie, T. J. Marks, M. A. Ratner, *Angew. Chem.-Int. Ed.* **2014**, *53*, 7456.

[39]    B. M. Savoie, A. Rao, A. A. Bakulin, S. Gelinas, B. Movaghar, R. H. Friend, T. J. Marks, M. A. Ratner, *J. Am. Chem. Soc.* **2014**, *136*, 2876.

[40]    B. M. Savoie, N. E. Jackson, L. X. Chen, T. J. Marks, M. A. Ratner, *Acc. Chem. Res.* **2014**, *47*, 3385.

[41]    L. Lu, L. Yu, *Adv. Mater.* **2014**, *26*, 4413.

[42]    V. D. Mihailetchi, L. J. A. Koster, P. W. M. Blom, C. Melzer, B. de Boer, J. K. J. van Duren, R. a. J. Janssen, *Adv. Funct. Mater.* **2005**, *15*, 795.

[43]    C. Melzer, E. J. Koop, V. D. Mihailetchi, P. W. M. Blom, *Adv. Funct. Mater.* **2004**, *14*, 865.

[44]    B. M. Savoie, K. L. Kohlstedt, N. E. Jackson, L. X. Chen, M. O. de la Cruz, G. C. Schatz, T. J. Marks, M. A. Ratner, *Proc. Natl. Acad. Sci.* **2014**, *111*, 10055.

[45]    D. Zhao, Q. Wu, Z. Cai, T. Zheng, W. Chen, J. Lu, L. Yu, *Chem. Mater.* **2016**, *28*, 1139.

[46]    A. Behranginia, P. Yasaei, A. K. Majee, V. K. Sangwan, F. Long, C. J. Foss, T. Foroozan, S. Fuladi, M. R. Hantehzadeh, R. Shahbazian-Yassar, M. C. Hersam, Z. Aksamija, A. Salehi-Khojin, *Small* **2017**, *13*, n/a.

[47]    L. K. J. Vandamme, R. Alabedra, M. Zommiti, *Solid-State Electron.* **1983**, *26*, 671.

[48]    T. G. M. Kleinpenning, F. Schurink, J. H. C. Van Der Veer, *Sol. Cells* **1984**, *12*, 363.

[49]    L. K. J. Vandamme, *IEEE Trans. Electron Devices* **1994**, *41*, 2176.

[50]    R. Dong, C. Bi, Q. Dong, F. Guo, Y. Yuan, Y. Fang, Z. Xiao, J. Huang, *Adv. Opt. Mater.* **2014**, *2*, 549.

[51]    Y. Yao, Y. Liang, V. Shrotriya, S. Xiao, L. Yu, Y. Yang, *Adv. Mater.* **2007**, *19*, 3979.

[52]    Y. Fang, F. Guo, Z. Xiao, J. Huang, *Adv. Opt. Mater.* **2014**, *2*, 348.

[53]    F. Guo, Z. Xiao, J. Huang, *Adv. Opt. Mater.* **2013**, *1*, 289.

[54]    A. Armin, R. D. Jansen-van Vuuren, N. Kopidakis, P. L. Burn, P. Meredith, *Nat. Commun.* **2015**, *6*, 6343.

[55]    A. T. Williams, P. Farrar, A. J. Gallant, D. Atkinson, C. Groves, *J. Mater. Chem. C* **2014**, *2*, 1742.

[56]    C. G. Shuttle, B. O'Regan, A. M. Ballantyne, J. Nelson, D. D. C. Bradley, J. de Mello, J. R. Durrant, *Appl. Phys. Lett.* **2008**, *92*, 093311.

[57]    R. Hamilton, C. G. Shuttle, B. O'Regan, T. C. Hammant, J. Nelson, J. R. Durrant, *J. Phys. Chem. Lett.* **2010**, *1*, 1432.

[58]    T. M. Clarke, C. Lungenschmied, J. Peet, N. Drolet, A. J. Mozer, *Adv. Energy Mater.* **2015**, *5*, 1401345.

[59]    S. Y. Kok, Z.-C. Hsieh, C.-H. Chou, S.-S. Yang, M.-K. Chuang, Y.-T. Lin, S. S. Yap, T. Y. Tou, F.-C. Chen, *Sci. Adv. Mater.* **2017**, *9*, 1435.

[60]    B. J. Leever, C. A. Bailey, T. J. Marks, M. C. Hersam, M. F. Durstock, *Adv. Energy Mater.* **2012**, *2*, 120.

[61]    F. Fabregat-Santiago, J. Bisquert, L. Cevey, P. Chen, M. Wang, S. M. Zakeeruddin, M. Graetzel, *J. Am. Chem. Soc.* **2009**, *131*, 558.

[62]    M. M. Rahman, N. C. D. Nath, J.-J. Lee, *Isr. J. Chem.* **2015**, *55*, 990.

[63]    F. Fabregat-Santiago, J. Bisquert, G. Garcia-Belmonte, G. Boschloo, A. Hagfeldt, *Sol. Energy Mater. Sol. Cells* **2005**, *87*, 117.

[64]    F. Fabregat-Santiago, G. Garcia-Belmonte, I. Mora-Seró, J. Bisquert, *Phys. Chem. Chem. Phys.* **2011**, *13*, 9083.

[65]    G. Garcia-Belmonte, A. Munar, E. M. Barea, J. Bisquert, I. Ugarte, R. Pacios, *Org. Electron.* **2008**, *9*, 847.

[66]    Y. Liang, Z. Xu, J. Xia, S.-T. Tsai, Y. Wu, G. Li, C. Ray, L. Yu, *Adv. Mater.* **2010**, *22*, E135.

[67]    S.-H. Liao, H.-J. Jhuo, Y.-S. Cheng, S.-A. Chen, *Adv. Mater.* **2013**, *25*, 4766.

[68]    S. H. Park, A. Roy, S. Beaupré, S. Cho, N. Coates, J. S. Moon, D. Moses, M. Leclerc, K. Lee, A. J. Heeger, *Nat. Photonics* **2009**, *3*, 297.

[69]    T. Kleinpenning, *J. Vac. Sci. Technol. -Vac. Surf. Films* **1985**, *3*, 176.

[70]    M. D. Perez, C. Borek, S. R. Forrest, M. E. Thompson, *J. Am. Chem. Soc.* **2009**, *131*, 9281.

[71]    Y. Sun, J. H. Seo, C. J. Takacs, J. Seifter, A. J. Heeger, *Adv. Mater.* **2011**, *23*, 1679.

[72]    S. Dongaonkar, M. A. Alam, *PV Analyzer*, **2014**.

[73]    J. D. Servaites, M. A. Ratner, T. J. Marks, *Energy Environ. Sci.* **2011**, *4*, 4410.

[74]    S.-H. Liao, H.-J. Jhuo, P.-N. Yeh, Y.-S. Cheng, Y.-L. Li, Y.-H. Lee, S. Sharma, S.-A. Chen, *Sci. Rep.* **2014**, *4*, 6813.

[75]    C. Cui, W.-Y. Wong, Y. Li, *Energy Environ. Sci.* **2014**, *7*, 2276.

[76]    I. P. Murray, S. J. Lou, L. J. Cote, S. Loser, C. J. Kadleck, T. Xu, J. M. Szarko, B. S. Rolczynski, J. E. Johns, J. Huang, L. Yu, L. X. Chen, T. J. Marks, M. C. Hersam, *J. Phys. Chem. Lett.* **2011**, *2*, 3006.

[77]    K. A. Luck, T. A. Shastry, S. Loser, G. Ogien, T. J. Marks, M. C. Hersam, *Phys. Chem. Chem. Phys.* **2013**, *15*, 20966.

[78]    S. Loser, B. Valle, K. A. Luck, C. K. Song, G. Ogien, M. C. Hersam, K. D. Singer, T. J. Marks, *Adv. Energy Mater.* **2014**, *4*, 1301938.

[79]    C. K. Song, K. A. Luck, N. Zhou, L. Zeng, H. M. Heitzer, E. F. Manley, S. Goldman, L. X. Chen, M. A. Ratner, M. J. Bedzyk, R. P. H. Chang, M. C. Hersam, T. J. Marks, *J. Am. Chem. Soc.* **2014**, *136*, 17762.

[80]    O. D. Jurchescu, B. H. Hamadani, H. D. Xiong, S. K. Park, S. Subramanian, N. M. Zimmerman, J. E. Anthony, T. N. Jackson, D. J. Gundlach, *Appl. Phys. Lett.* **2008**, *92*, 132103.

[81]    C. Bonavolontà, C. Albonetti, M. Barra, M. Valentino, *J. Appl. Phys.* **2011**, *110*, 093716.

[82]    Y. Xu, T. Minari, K. Tsukagoshi, J. Chroboczek, F. Balestra, G. Ghibaudo, *Solid-State Electron.* **2011**, *61*, 106.

[83]    L. Ke, S. B. Dolmanan, C. Vijila, S. J. Chua, Y. H. Han, T. Mei, *IEEE Trans. Electron Devices* **2010**, *57*, 385.

[84]    H. Kang, L. Jagannathan, V. Subramanian, *Appl. Phys. Lett.* **2011**, *99*, 062106.

[85]    G. Ferrari, D. Natali, M. Sampietro, F. P. Wenzl, U. Scherf, C. Schmitt, R. Güntner, G. Leising, *Org. Electron.* **2002**, *3*, 33.

[86]    L. Li, Y. Shen, J. C. Campbell, *Sol. Energy Mater. Sol. Cells* **2014**, *130*, 151.

[87]    H. Katsu, Y. Kawasugi, R. Yamada, H. Tada, in *2011 21st Int. Conf. Noise Fluct. ICNF*, **2011**, pp. 77–79.

[88]    G. Landi, C. Barone, A. D. Sio, S. Pagano, H. C. Neitzert, *Appl. Phys. Lett.* **2013**, *102*, 223902.

[89]    K. Kaku, A. T. Williams, B. G. Mendis, C. Groves, *J. Mater. Chem. C* **2015**, *3*, 6077.

[90]    M. Bag, N. S. Vidhyadhiraja, K. S. Narayan, *Appl. Phys. Lett.* **2012**, *101*, 043903.

[91]    F. N. Hooge, *Physica* **1972**, *60*, 130.

[92]    E. Burstein, A. L. McWhorter, *Semiconductor Surface Physics*, Literary Licensing, LLC, **2012**.

[93]    V. K. Sangwan, H. N. Arnold, D. Jariwala, T. J. Marks, L. J. Lauhon, M. C. Hersam, *Nano Lett.* **2013**, *13*, 4351.

[94]    C. Y. Han, L. X. Qian, C. H. Leung, C. M. Che, P. T. Lai, *J. Appl. Phys.* **2013**, *114*, 044503.

[95]    H. D. Xiong, B. Jun, D. M. Fleetwood, R. D. Schrimpf, J. R. Schwank, *IEEE Trans. Nucl. Sci.* **2004**, *51*, 3238.

[96]    R. W. Bené, G. S. Lee, N. I. Cho, *Thin Solid Films* **1985**, *129*, 195.

[97]    F. Hooge, T. Kleinpenning, L. Vandamme, *Rep. Prog. Phys.* **1981**, *44*, 479.

[98]    T. Kleinpenning, *Phys. B C* **1987**, *145*, 190.

[99]    T. Kleinpenning, *Phys. B C* **1986**, *142*, 229.

[100]    R. Macku, P. Koktavy, *Phys. Status Solidi A* **2010**, *207*, 2387.

[101]    G. Landi, C. Barone, C. Mauro, H. C. Neitzert, S. Pagano, *Sci. Rep.* **2016**, *6*, 29685.

[102]    A. Mehrabian, R. Hoheisel, D. J. Nagel, S. R. Messenger, S. Maximenko, *2014 IEEE 40th Photovolt. Spec. Conf. Pvsc* **2014**, 1884.

[103]    M. R. S. Abouzari, F. Berkemeier, G. Schmitz, D. Wilmer, *Solid State Ion.* **2009**, *180*, 922.

[104]    H. M. Heitzer, T. J. Marks, M. A. Ratner, *J. Am. Chem. Soc.* **2015**, *137*, 7189.

[105]    G. Garcia-Belmonte, P. P. Boix, J. Bisquert, M. Sessolo, H. J. Bolink, *Sol. Energy Mater. Sol. Cells* **2010**, *94*, 366.

[106]    J. Bisquert, *Phys. Chem. Chem. Phys.* **2003**, *5*, 5360.

[107]    J. van de Lagemaat, N.-G. Park, A. J. Frank, *J. Phys. Chem. B* **2000**, *104*, 2044.

[108]    F. Fabregat-Santiago, I. Mora-Seró, G. Garcia-Belmonte, J. Bisquert, *J. Phys. Chem. B* **2003**, *107*, 758.

[109]    J. Idigoras, R. Tena-Zaera, J. A. Anta, *Phys. Chem. Chem. Phys.* **2014**, *16*, 21513.

[110]    J. Bisquert, *Phys. Chem. Chem. Phys.* **2011**, *13*, 4679.

[111]    F. Fabregat-Santiago, G. Garcia-Belmonte, J. Bisquert, A. Zaban, P. Salvador, *J. Phys. Chem. B* **2002**, *106*, 334.

[112]    T. González, in *2015 Int. Conf. Noise Fluct. ICNF*, **2015**, pp. 1–6.

[113]    X. N. ZHANG, A. VAN DER ZIEL, in *Noise Phys. Syst. 1f Noise 1985* (Eds.: A. D'AMICO, P. MAZZETTI), Elsevier, Amsterdam, **1986**, pp. 485–487.

[114]    J. Vanek, J. Dolensky, Z. Chobola, M. Lunak, A. Poruba, *Int. J. Photoenergy* **2012**, 324853.

[115]    R. Jones, *J. Opt. Soc. Am.* **1960**, *50*, 1058.

[116]    G. Li, V. Shrotriya, J. Huang, Y. Yao, T. Moriarty, K. Emery, Y. Yang, *Nat. Mater.* **2005**, *4*, 864.

[117]    H. Yan, S. Swaraj, C. Wang, I. Hwang, N. C. Greenham, C. Groves, H. Ade, C. R. McNeill, *Adv. Funct. Mater.* **2010**, *20*, 4329.

[118]    P. E. Keivanidis, V. Kamm, W. Zhang, G. Floudas, F. Laquai, I. McCulloch, D. D. C. Bradley, J. Nelson, *Adv. Funct. Mater.* **2012**, *22*, 2318.

[119]    T. M. Clarke, C. Lungenschmied, J. Peet, N. Drolet, A. J. Mozer, *J. Phys. Chem. C* **2015**, *119*, 7016.

[120]    Y. J. Kim, S. Ahn, D. H. Wang, C. E. Park, *Sci. Rep.* **2015**, *5*, 18024.

[121]    N. D. Eastham, A. S. Dudnik, T. J. Aldrich, E. F. Manley, T. J. Fauvell, P. E. Hartnett, M. R. Wasielewski, L. X. Chen, F. S. Melkonyan, A. Facchetti, R. P. H. Chang, T. J. Marks, *Chem. Mater.* **2017**, *29*, 4432.

[122]    R. Sarpeshkar, T. Delbruck, C. A. Mead, *IEEE Circuits Devices Mag.* **1993**, *9*, 23.

[123]    J. Vanek, Z. Chobola, V. Brzokoupil, J. Kazelle, in *Noise Devices Circuits III* (Eds.: A.A. Balandin, F. Danneville, M.J. Deen, D.M. Fleetwood), SPIE-Int Soc Optical Engineering, Bellingham, **2005**, pp. 86–93.

[124]    A. Vanderziel, *Solid-State Electron.* **1982**, *25*, 141.

[125]    X. Wu, J. B. Anderson, A. van der Ziel, *IEEE Trans. Electron Devices* **1987**, *34*, 1971.

[126]   I. Etxebarria, A. Guerrero, J. Albero, G. Garcia-Belmonte, E. Palomares, R. Pacios, *Org. Electron.* **2014**, *15*, 2756.

[127]   S. R. Cowan, A. Roy, A. J. Heeger, *Phys. Rev. B* **2010**, *82*, 245207.

[128]   S. R. Cowan, W. L. Leong, N. Banerji, G. Dennler, A. J. Heeger, *Adv. Funct. Mater.* **2011**, *21*, 3083.

[129]   D. Neher, J. Kniepert, A. Elimelech, L. J. A. Koster, *Sci. Rep.* **2016**, *6*, 24861.

[130]   R. Noriega, J. Rivnay, K. Vandewal, F. P. V. Koch, N. Stingelin, P. Smith, M. F. Toney, A. Salleo, *Nat. Mater.* **2013**, *12*, 1038.

[131]   Y. Li, *Open Photovoltaics Analysis Platform (OPVAP)*, **2011**.

[132]   H. N. Arnold, V. K. Sangwan, S. W. Schmucker, C. D. Cress, K. A. Luck, A. L. Friedman, J. T. Robinson, T. J. Marks, M. C. Hersam, *Appl. Phys. Lett.* **2016**, *108*, 073108.

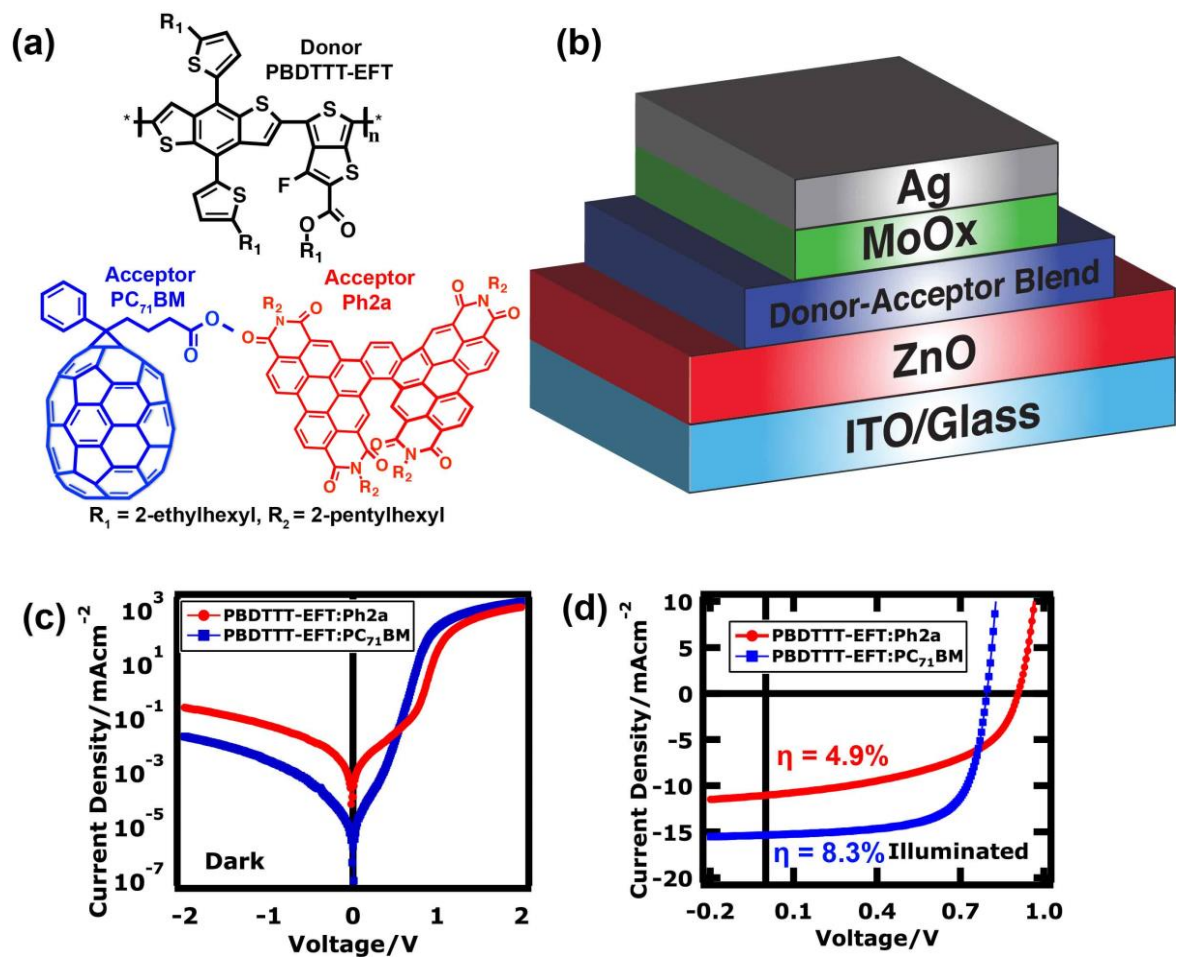

**Figure 1.** (a) Chemical structures of PBDTTT-EFT, $PC_{71}BM$, and Ph2a. (b) Inverted organic photovoltaic (OPV) device architecture. (c) Representative dark current-voltage data for PBDTTT-EFT:$PC_{71}BM$ and PBDTTT-EFT:Ph2a solar cells. (d) Representative illuminated current-voltage data for PBDTTT-EFT:$PC_{71}BM$ and PBDTTT-EFT:Ph2a solar cells.

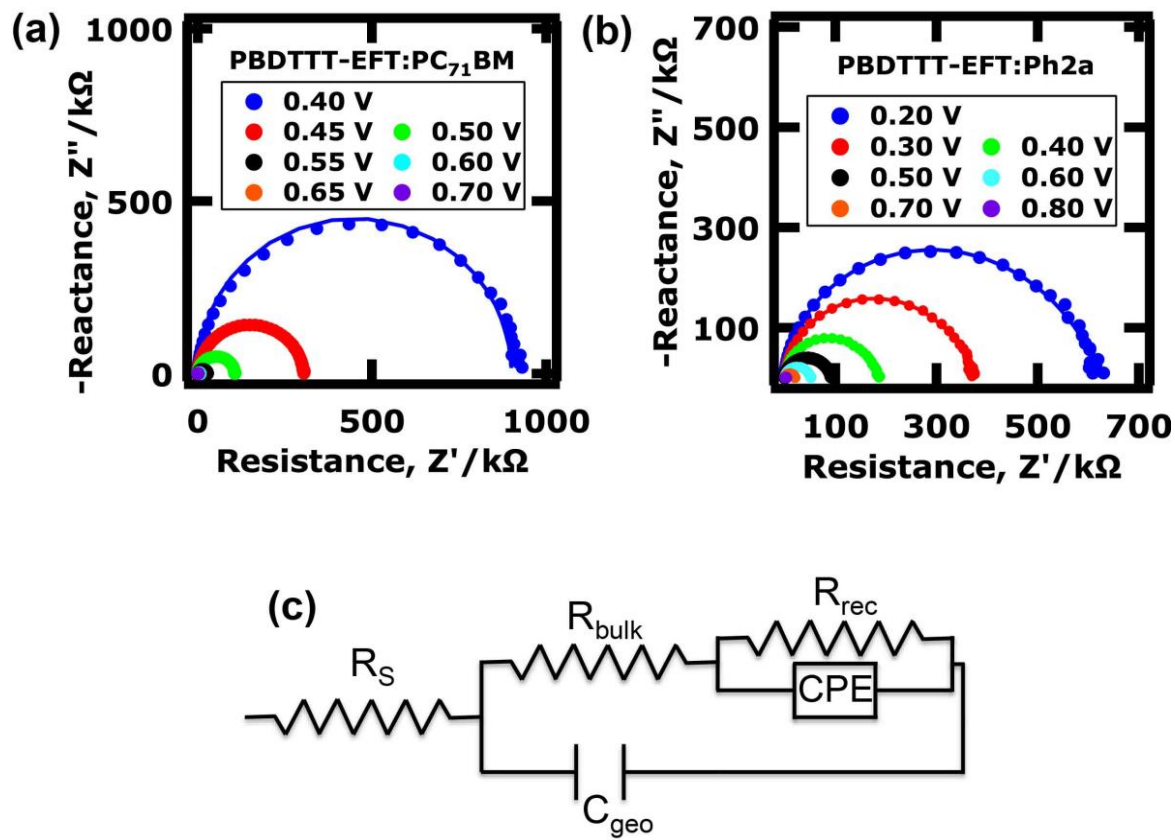

**Figure 2.** (a) Impedance response of a PBDTTT-EFT:PC$_{71}$BM solar cell for different biases in the dark. (b) Impedance response of a PBDTTT-EFT:Ph2a solar cell for different biases in the dark. (c) Equivalent circuit model for dark PBDTTT-EFT:PC$_{71}$BM and PBDTTT-EFT:Ph2a OPV impedance data (R$_s$ – series resistance, R$_{bulk}$ – bulk resistance, C$_{geo}$ – geometrical capacitance, R$_{rec}$ – recombination resistance, CPE – constant phase element representing chemical capacitance).

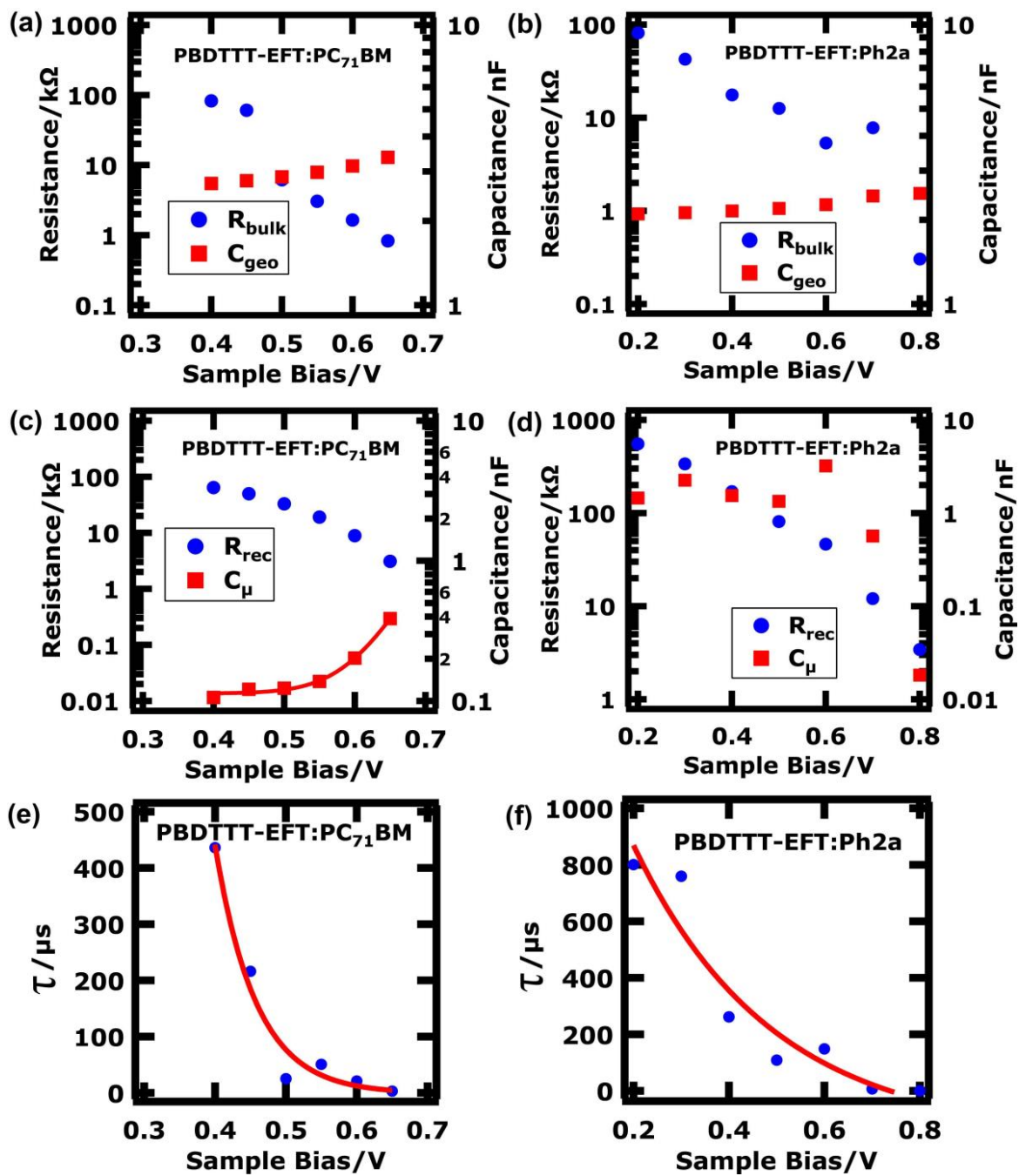

**Figure 3.** (a) Bulk resistance ($R_{bulk}$) and geometrical capacitance ($C_{geo}$) impedance circuit elements in the dark for a PBDTTT-EFT:$PC_{71}$BM OPV. (b) Bulk resistance ($R_{bulk}$) and geometrical capacitance ($C_{geo}$) impedance circuit elements in the dark for a PBDTTT-EFT:Ph2a OPV. (c) Recombination resistance ($R_{rec}$) and chemical capacitance ($C_\mu$) of a PBDTTT-EFT:$PC_{71}$BM OPV in the dark at varied sample bias. The line is a power law fit to the chemical capacitance data. (d) Recombination resistance ($R_{rec}$) and chemical capacitance ($C_\mu$) of a PBDTTT-EFT:Ph2a OPV in the dark under varied sample bias. (e) Average charge carrier lifetime ($\tau$) of a PBDTTT-EFT:$PC_{71}$BM OPV in the dark at varied sample bias. The line is an exponential fit to the data. (f) Average charge carrier lifetime ($\tau$) of a PBDTTT-EFT:Ph2a OPV in the dark at varied sample bias. The line is an exponential fit to the data.

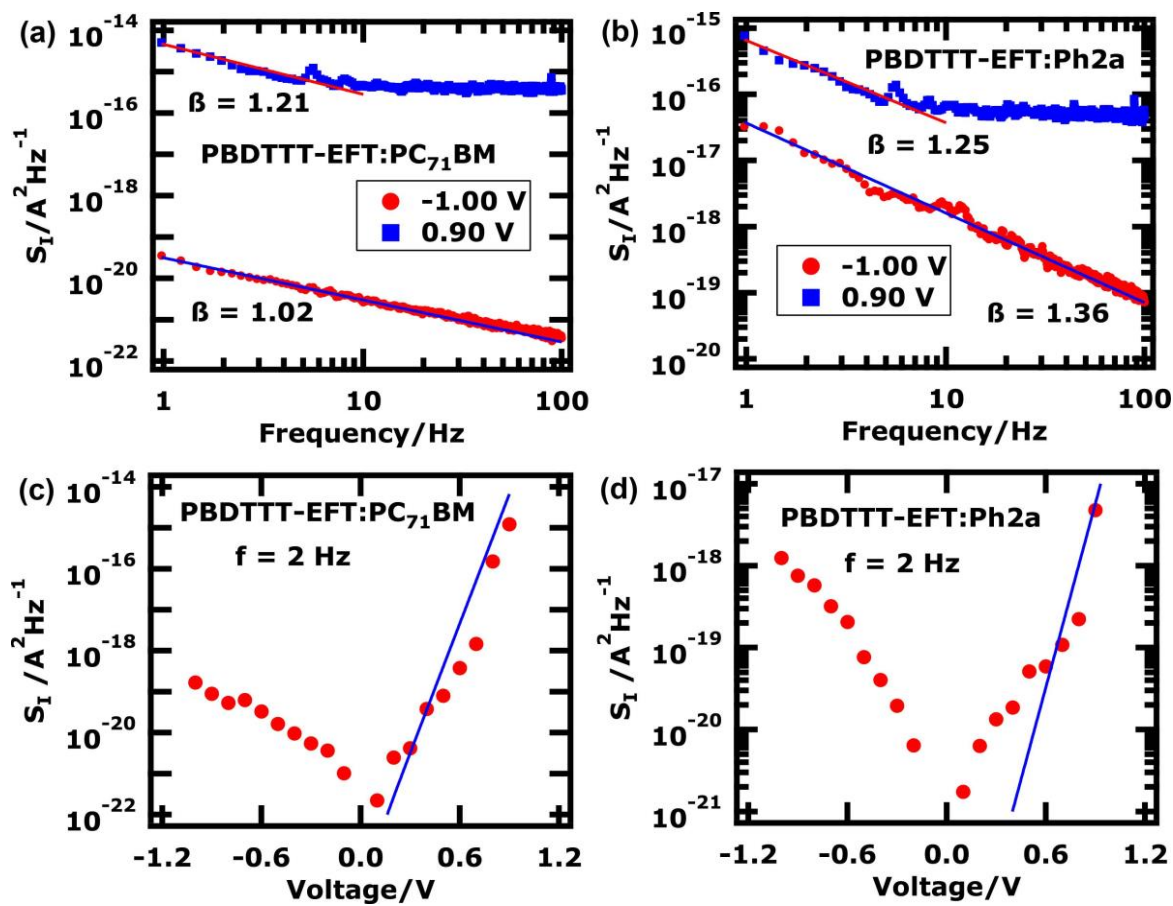

**Figure 4.** (a) Noise spectral density ($S_I$) versus frequency of a PBDTTT-EFT:PC$_{71}$BM solar cell in the dark, showing $1/f^\beta$ behavior at V = -1.00 V with $\beta = 1.02 \pm 0.005$ and a $1/f^\beta$ noise spectrum at $V = 0.90$ V for $f < 10$ Hz with $\beta = 1.21 \pm 0.05$. (b) Noise spectral density ($S_I$) versus frequency of a PBDTTT-EFT:Ph2a OPV in the dark, showing $1/f^\beta$ behavior at $V = -1.00$ V with $\beta = 1.36 \pm 0.015$ and a $1/f^\beta$ spectrum at V = 0.90 V for $f < 10$ Hz with $\beta = 1.25 \pm 0.05$. (c) Noise spectral density ($S_I$) at $f = 2$ Hz versus voltage of a PBDTTT-EFT:PC$_{71}$BM OPV in the dark. The lines are fit to the Kleinpenning model (eq. 3) under forward bias. (d) Noise spectral density ($S_I$) at $f = 2$ Hz versus voltage of a PBDTTT-EFT:Ph2a OPV in the dark. The lines are fit to the Kleinpenning model (eq. 3) under forward bias.

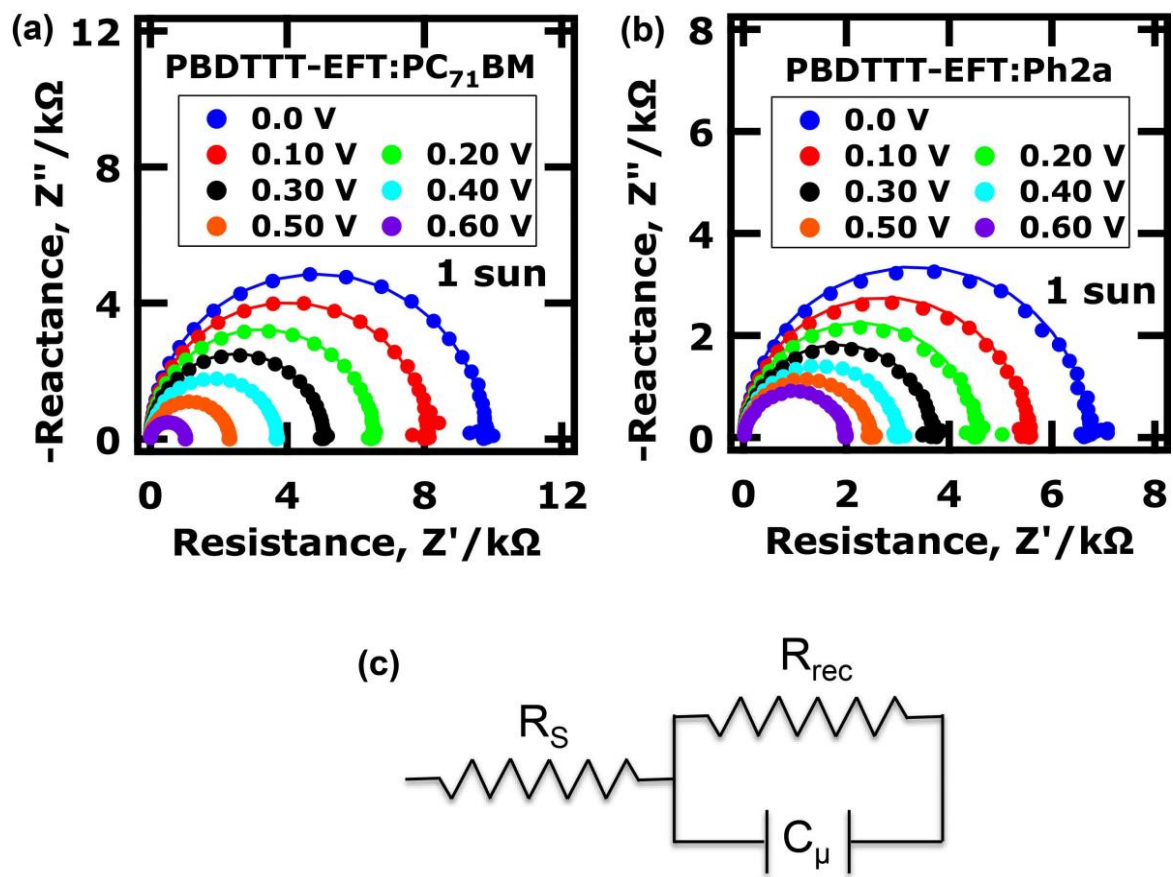

**Figure 5.** (a) Impedance response of a PBDTTT-EFT:PC$_{71}$BM solar cell under 1 sun illumination. (b) Impedance response of a PBDTTT-EFT:Ph2a solar cell under 1 sun illumination. (c) Equivalent circuit model for illuminated PBDTTT-EFT:PC$_{71}$BM and PBDTTT-EFT:Ph2a OPV impedance data (R$_s$ – series resistance, R$_{rec}$ – recombination resistance, C$_\mu$ – chemical capacitance).

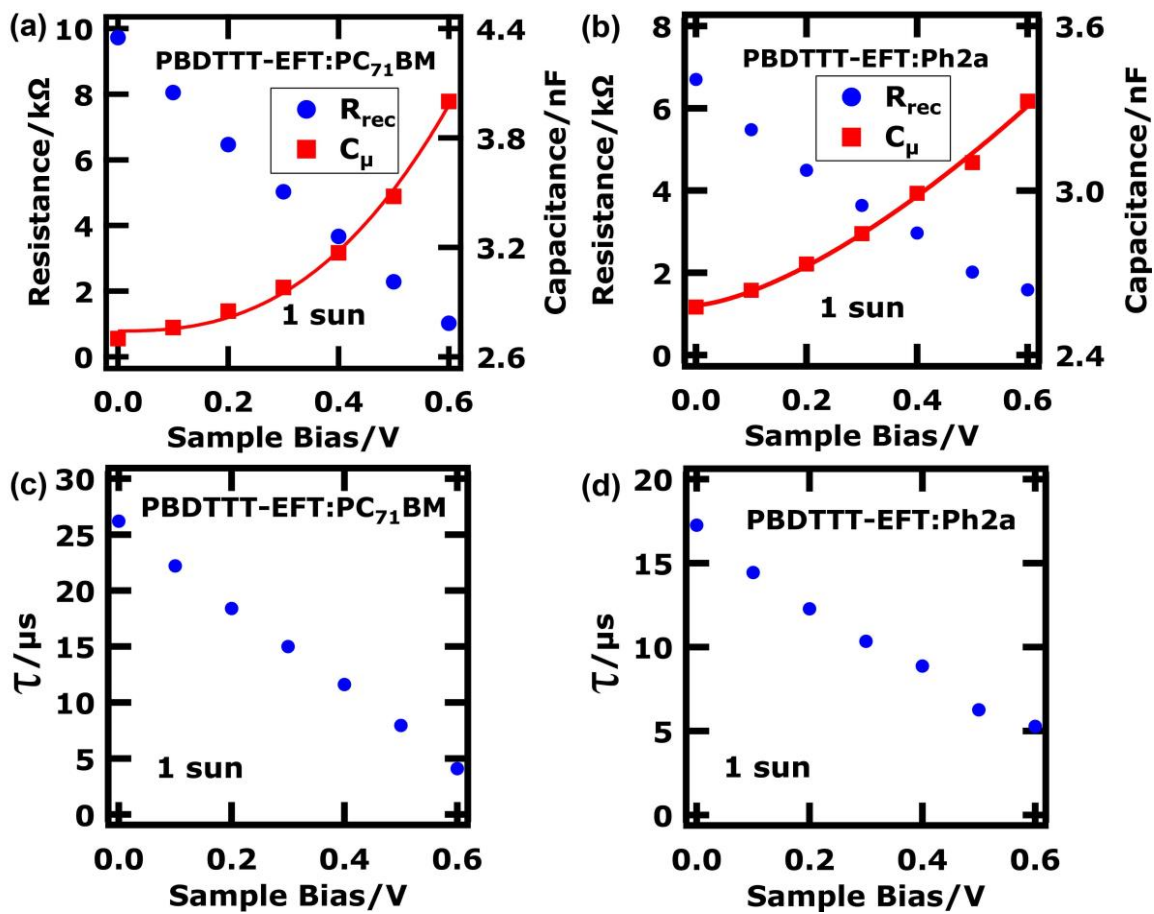

**Figure 6.** (a) Recombination resistance (R_rec) and chemical capacitance (C_μ) of a PBDTTT-EFT:PC_71BM OPV under 1 sun illumination at varying sample bias. The line is a power law fit to the chemical capacitance data. (b) Recombination resistance (R_rec) and chemical capacitance (C_μ) of a PBDTTT-EFT:Ph2a OPV under 1 sun illumination under varying sample bias. The line is a power law fit to the chemical capacitance data. (c) Average charge carrier lifetime ($\tau$) of a PBDTTT-EFT:PC_71BM OPV under 1 sun illumination at different sample biases. (d) Average charge carrier lifetime ($\tau$) of a PBDTTT-EFT:Ph2a OPV in the dark at different sample bias.

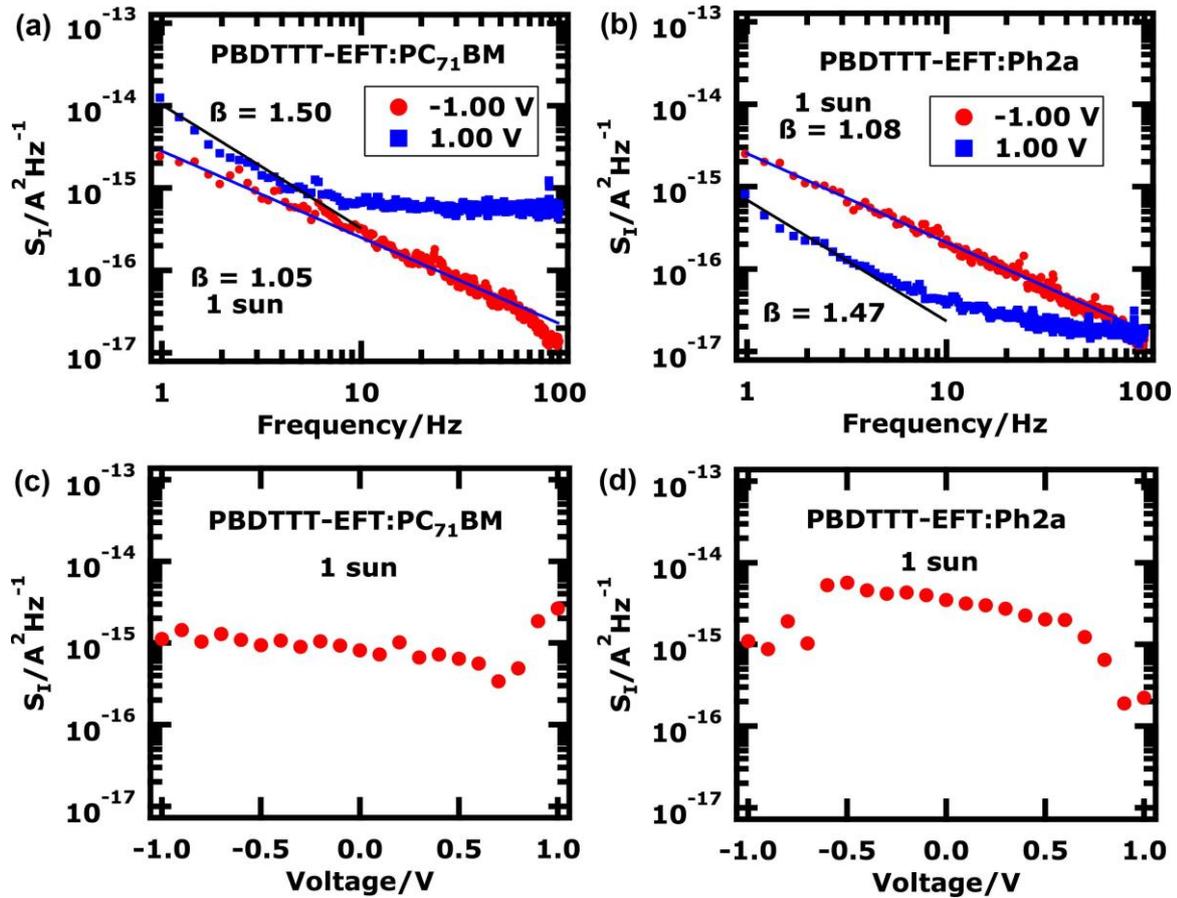

**Figure 7.** Noise spectral density ($S_I$) versus frequency of a PBDTTT-EFT:PC$_{71}$BM solar cell under 1 sun illumination, showing $1/f^\beta$ behavior at $V$ = -1.00 V with $\beta$ = 1.05 ± 0.01 and a ~$1/f^\beta$ noise spectrum at $V$ = 1.00 V for $f$ < 10 Hz with $\beta$ = 1.50 ± 0.07. (b) Noise spectral density ($S_I$) versus frequency of a PBDTTT-EFT:Ph2a solar cell under 1 sun illumination, showing $1/f^\beta$ behavior at $V$ = -1.00 V with $\beta$ = 1.08 ± 0.006 and a ~$1/f^\beta$ noise spectrum at $V$ = 1.00 V for $f$ < 10 Hz with $\beta$ = 1.47 ± 0.06. (c) Noise spectral density ($S_I$) at $f$ = 2 Hz versus voltage of a PBDTTT-EFT:PC$_{71}$BM solar cell under 1 sun illumination. (d) Noise spectral density ($S_I$) at $f$ = 2 Hz versus voltage of a PBDTTT-EFT:Ph2a solar cell under 1 sun illumination.

WILEY-VCH

**TABLE OF CONTENTS**

**Low-frequency electronic noise is measured** in polymer solar cells with fullerene and non-fullerene acceptors. Charge carrier lifetimes deduced from impedance spectroscopy enable the noise data to be fit to the Kleinpenning model. The results establish that low-frequency noise elucidates charge recombination processes that limit power conversion efficiency. This correlated analytical tool provides quantitative guidance to the optimization of emerging photovoltaic materials.

**Keywords:**
$1/f$ noise, organic photovoltaic, PBDTTT-EFT, PC$_{71}$BM, alternate acceptor

Kyle A. Luck, Vinod K. Sangwan, Patrick E. Hartnett, Heather N. Arnold, Michael R. Wasielewski, Tobin J. Marks, and Mark C. Hersam*

**Correlated In-Situ Low-Frequency Noise and Impedance Spectroscopy Reveal Recombination Dynamics in Organic Solar Cells using Fullerene and Non-Fullerene Acceptors**

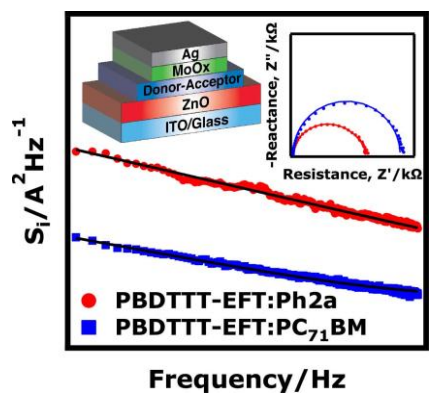